\documentclass[journal=jceaax,manuscript=article]{achemso}

\usepackage[version=3]{mhchem} 
\usepackage{enumerate}
\usepackage[shortlabels]{enumitem}

 \usepackage{setspace}
 \usepackage{natbib}

\usepackage{amsmath}
\usepackage{upgreek}
\usepackage{graphicx}
\usepackage{chemfig}
\usepackage{caption}
\usepackage{textcomp}
\usepackage{underscore}
\usepackage{gensymb}
\usepackage{dcolumn}
\usepackage{siunitx}
\usepackage{multirow}
\usepackage{booktabs}
\usepackage{tabularx}
\usepackage{adjustbox}
\usepackage{braket}
\usepackage{color, colortbl}
\usepackage{algorithm}
\usepackage{arevmath}     
\usepackage{mathtools}
\usepackage[noend]{algpseudocode}
\usepackage{algorithm,algpseudocode}
\definecolor{LightCyan}{rgb}{0.88,1,1}
\definecolor{Gray}{gray}{0.9}
\usepackage[first=0,last=9]{lcg}

\usepackage{dsfont}
\usepackage{threeparttable} 

\setcitestyle{number} 

\author{Rodrigo A. Mendes}
\affiliation[University of Florida]{Quantum Theory Project, University of Florida, Gainesville, FL 32611, USA.}

\author{Zachary W. Windom}
\affiliation[University of Florida]{Quantum Theory Project, University of Florida, Gainesville, FL 32611, USA.}

\author{Ajith Perera}
\affiliation[University of Florida]{Quantum Theory Project, University of Florida, Gainesville, FL 32611, USA.}

\author{Roberto L. A. Haiduke}
\affiliation[USP]{S{\~a}o Carlos Institute of Chemistry, University of S{\~a}o Paulo, Av. Trabalhador S{\~a}o-Carlense, 400, 13566-590 S{\~a}o Carlos, SP, Brazil.}

\author{Rodney J. Bartlett}
\affiliation[University of Florida]{Quantum Theory Project, University of Florida, Gainesville, FL 32611, USA.}
\email{bartlett@qtp.ufl.edu}

\title{A note on the accuracy of spin-densities from Kohn-Sham Density Functional Theory}

\begin{document}

\newpage

\begin{abstract}
Quantifying the magnetic properties of open-shell molecules is a common task in chemistry and is increasingly performed \emph{in silico} using Kohn-Sham density functional theory (KS-DFT). Previous work demonstrates that the predictive accuracy of a few functionals for one such property - hyperfine coupling constants (HFCCs) - is possible, implying that such approximations must yield accurate spin-densities at the nucleus. However, the ability of such functionals to globally predict accurate spin-densities of comparable quality to rigorous \emph{ab initio} coupled cluster theory, for example, is dubious, despite this being a fundamental quantity for KS-DFT. This work intends to explore the matter by evaluating moments of the spin-density, $\braket{r^n}= \int \rho(r)r^n d\tau$, $n=-2,\cdots,2,3$, for second-, third-, and fourth-row atoms to compare various KS-DFT functionals against coupled cluster theory with single and double excitations (CCSD). Our results broadly indicate that the tested functionals experience significant deviations with respect to CCSD spin-densities in regions up to 1 Bohr away from the nuclei, with most errors occurring in the immediate vicinity of the nucleus. We find evidence of extreme errors by some functionals for individual $\alpha$/$\beta$ spin-densities, although several ultimately end up benefiting from significant error cancellation once the corresponding global spin-density is formed. Nevertheless, a comparison between CAM-B3LYP and the Quantum Theory Project (QTP)-family of DFT functionals based on Correlated Orbital Theory conditions across all error metrics demonstrates that QTP00 more accurately reproduces the global spin-density as well as HFCCs, generally offering results in better agreement with CCSD. In line with previous work, we also corroborate the success of PBE0 and the TPSS-family of functionals for HFCCs, further finding that both approximations generally yield spin-densities that are amongst the best.
\end{abstract}  

\newpage

\section{Introduction}
The need for an accurate assessment of magnetic properties for open-shell systems has become increasingly relevant for technological applications. This encourages the development of robust theory to support ongoing scientific discovery $\emph{in silico}$. To this end, early efforts using wavefunction theory focused on small radicals, with some of the most impressive results coming from coupled cluster theory.\cite{perera1994theoretical,perera1994coupled,perera1996electron} Although these developments have been incorporated into software designed to take advantage of massive parallel architectures, as shown by programs like ACESIII,\cite{verma2013massively,perera2017benchmark} the computational scaling of many-body theory instigates a soft limit of around $1000$ basis functions in a calculation if no approximations are used. This has encouraged the development of techniques designed to reduce calculation costs; see, for example, the idea behind frozen natural orbitals. On the other hand, Kohn-Sham density functional theory (KS-DFT)\cite{kohn1965self} is an inherently cheap alternative wherein a density functional approximation (DFA) is designed to consider both exchange and correlation effects within an effective one-particle framework. Today, there are many DFA choices, which are becoming increasingly difficult to differentiate.\cite{rappoport2009functional} 

After the success of the Local Density Approximation (LDA),\cite{kohn1965self,ceperley1980ground,perdew1992accurate} attempts to improve the functional design soon began. This has led to the several flavors of DFT functionals available today, such as Generalized Gradient Approximations (GGAs),\cite{perdew1996generalized} which include electron density gradient information, Meta-GGAs, which incorporate the Laplacian of the density pertinent to the kinetic energy density,\cite{tao2003climbing} and Hybrid (HYB) functionals, which include some percentage of non-local, Hartree-Fock (HF)-like, exchange.\cite{becke1993new} Nevertheless, it can be generally asserted that the majority of currently available density functionals have either been (over-) parameterized with a particular subset of properties in mind,\cite{zhao2008m06} or focus too strongly on model systems such as the uniform electron gas, while largely neglecting conditions that would be rigorously satisfied by a truly $\emph{ab initio}$ DFT functional.\cite{bartlett2019adventures} Another noticeable failure of traditional DFAs is that they lack a rigorous route to systematically converge toward the full Configuration Interaction (FCI) limit, which should be the target for any electronic structure method.  

Although it can be argued that improvements have been made over the years,\cite{oliphant1994systematic} issues associated with the ``Devil's Triangle of KS-DFT"\cite{bartlett2019adventures,mendes2021devil} continue to persist in traditional DFT functionals. Notably, problems associated with the correct one-particle spectrum, self-interaction error, and integer discontinuity\cite{mori2014derivative,perdew1982density} continue to plague modern functionals' ability to adequately capture charge transfer,\cite{baer2010tuned,dreuw2003long} Rydberg states,\cite{mendes2021devil,gudmundsdottir2013self} and transition states.\cite{jin2016qtp,Mendes_2026} When left unaddressed, these issues limit the accuracy and general applicability of a density functional.\cite{bartlett2017power}   

Early work in Bartlett's group pursued \emph{ab initio} DFT,\cite{bartlett2005ab} which focused on systematically converging to the FCI result by using the Optimized Effective Potential (OEP) strategy.\cite{talman1976optimized,hirata2001can} These initial OEP efforts targeted the exact local exchange potential. Encouraged by the results,\cite{hirata2002time} custom correlation functionals and potentials were derived from coupled cluster (CC) and many-body perturbation theory (MBPT) results. It is important to emphasize that  the OEP procedure constructs potentials that are completely devoid of any self-interaction error, which is arguably the most significant source of error that arises from the ``Devil's Triangle of KS-DFT." Using arguments from a Correlated Orbital Theory (COT) designed to map CC theory into an effective one-particle framework,\cite{bartlett2009towards} it was shown that reducing the self-interaction error improves orbital eigenvalues to largely coincide with principal ionization potentials (IPs) - akin to a Koopmans' theorem for DFT -  thereby establishing a profound relationship between the sides of the ``Devil's Triangle." Unsurprisingly, these $\emph{ab initio}$  methods based on COT demonstrated appreciable improvements in several instances as compared to traditional DFT approximations.\cite{bartlett2005ab} However, these improvements come at a sizable cost. For example, the correlation potentials generated from the OEP procedure require inversion of the correlated MBPT or CC density matrices. 

In the context of traditional KS-DFT, the main takeaway from this body of work is clear: to improve a traditional functionals' performance, parameterize it in such a way that the orbital eigenvalues correspond to principal IPs,\cite{chong2002interpretation} as required by COT. The result of this parameterization constraint is an accurate KS potential, leading to meaningful orbitals and improved densities. As alluded to earlier, doing this inherently mitigates the self-interaction error that conveniently alleviates issues attributable to the ``Devil's Triangle of KS-DFT." To verify these claims, the CAM-B3LYP functional\cite{yanai2004new} was re-parameterized in different ways to satisfy the COT conditions, leading to the QTP00,\cite{verma2014increasing} QTP01,\cite{jin2016qtp} QTP02,\cite{haiduke2018qtp2} and LC-QTP\cite{haiduke2018qtp2} functionals. It is now well-documented that the QTP-line of functionals generally improves several properties\cite{Kim_2025,Mendes_2025} and excitation spectra predictions as compared to other DFT functionals in common use today.\cite{windom2022examining,mendes2021devil,mendes2021performance,verma2016increasing}

In this regard, a recent benchmark published by the Bartlett's group compares DFT hyperfine coupling constant predictions for several small, medium, and large organic radicals - as well as transition metal complexes - against CCSD and CCSD(T).\cite{windom2022benchmarking} This work emphasized the surprising agreement that a majority of DFT functionals show with CC, where more than half of the functionals in the study had average errors $\leq 10$ MHz. Some noticeably outperforming functionals include OLYP (GGA), TPSSh (Hybrid of Meta-GGA, MHG), and PBE0 (HYB). On the other hand, the SCAN functional was routinely among the worst performers in spite of its parameterization constraints to satisfy certain exact conditions. 

It was also tangentially noticed that QTP01 and QTP02 exhibited the overall best performance. This outcome seemingly conflicts with how these functionals were parameterized, where heavier emphasis was devoted to valence orbitals. In addition to this enigma, the two functionals specifically parameterized for core electrons, the global hybrid QTP17 and QTP00, performed the worst among the QTP functionals. Although outer-electron contributions to hyperfine coupling constants (HFCCs) are increasingly relevant in transition metals,\cite{abragam1955hyperfine,abragam2012electron} the above trends were also seen in the organic radical test sets containing both small and large molecules. It was proposed that the poor performance of the core-electron parameterized QTP17 and QTP00 functionals was due to the usage of too much HF exchange (in the short-range for QTP00), which leads to results that are similar to those of HF itself. 

Of course, a particular functional's success or failure at predicting HFCCs ultimately depends on the spin-density accuracy at one particular point in space: the nucleus. However, this previous work does not lend any insight into the global spin-density behavior of any functional. In this context, a more complete analysis of global spin-densities is required to complement the findings of the aforementioned work.\cite{windom2022benchmarking} Thus, this study serves as a follow-up where we intend to survey the global spin-density of atoms and their moments, as predicted by various DFT functionals and CCSD. 

Others have pursued similar objectives for transition metal complexes by visualizing spin-density distributions at particular sites,\cite{remenyi2007density,remenyi2006density,spada2022spin} studying the radial distribution function of the spin-density from UHF/ROHF/UKS(and CCSD/CCSD(T) to a limited extent),\cite{munzarova2000mechanisms,datta2015communication}  or decomposing the isotropic HFCCs into contributions from individual Kohn-Sham molecular orbitals.\cite{hedegard2013validating} The current work studies atomic radicals up to the fourth row, using a myriad of DFT functionals that include popular members of the GGA, MHG, HYB, and range-separated hybrid (RSH) families. This facilitates a robust comparison of KS-DFT spin-density moments to values from CCSD and, as such, the current work complements the existing understanding on the matter.

In this case, the moments are written in terms of the distance to the nucleus, $r$, and spin-densities, $\rho^{SD}(r)=\rho^{\alpha}(r)- \rho^{\beta}(r)$, that is,
\begin{equation}
\braket{r^n}= \int \rho^{SD}(r) r^n d\tau = \int_0^{\infty} 4\pi r^2\rho^{SD}(r) r^n dr
\end{equation} for $n=-2,\cdots,2,3$. This range of $n$ surveys particular regions of space, where  $n<0$ emphasizes the regions around the core and $n>0$ emphasizes the valence space. Thus, the usefulness behind calculating several moment integrals can be reinforced with an example:  as $r\rightarrow 0$ the magnitude of $r^{-2}$ and $r^{-1}$ increases,  which happens more rapidly with the former. Hence, even slight deviations in the predicted spin-density will be appreciably amplified, which facilitates a more robust examination of density functional performance.  
 
Additionally, the current work builds off a study by Ranasinghe et al. on the global electron density for closed-shell atoms.\cite{ranasinghe2017note} Similar studies have also been published that seek a more thorough understanding of the electron density as predicted by an assortment of KS-DFT functionals. Consequently, commentary that explains the rationale behind studying the KS-DFT spin-density, while largely ignoring the density that has historically been the canonical variable of KS-DFT, is merited.   

In this context, the seminal work by Kohn and Sham\cite{kohn1965self} insists that the interacting and non-interacting model systems share the same electron density, despite not having equivalent wavefunctions. This remains true independently of the reference choice (i.e., restricted or unrestricted) used to define the non-interacting model system. In contrast, the non-interacting model system shares an equivalent spin-density with the interacting model system only if the former uses an unrestricted reference.\cite{jacob2012spin} If we only consider closed shells where the spin-density is null by definition, then equivalence is established between the restricted and unrestricted reference choices, and the theorem by Kohn and Sham is recovered. Regardless, these considerations imply that the unrestricted formulation of KS-DFT and, by extension, the spin-densities, are fundamental objects worthy of investigation.\cite{saue2002four}

In summary, this work intends to serve as an initial assessment of global spin-density behavior as predicted by several DFAs and CCSD. Because a consistent way to assess KS-DFT functional performance in predicting accurate spin-densities is inherently opaque since the density functionals are approximate, it is very difficult to extract general conclusions that can provide meaningful details on a KS-DFT functionals' reliability for arbitrary systems.\cite{reiher2007definition} Thus, a rigorous analysis of KS-DFT spin-densities as compared against CCSD is warranted.  We anticipate that this work will further quantify and expose current DFA shortcomings, in the hope that future exchange-correlation functionals can be more rigorously designed in accordance with COT.

\section{Computational Details}

All DFT calculations are performed in NWChem,\cite{nwchem} whereas CCSD, MBPT\textit{n} (\textit{n} = 2, 3, and 4), and moment calculations were carried out inside ACES2.\cite{perera2020advanced} Software was written to extract converged self-consistent field (SCF) vectors from NWChem output, which interfaced directly to the moment integral calculation subroutines recently constructed in ACES2. The radial distribution function and the spin-density as a function of $r$ are also accessible in these new subroutines. Integration of the $\braket{r^n}$  integrals is carried out numerically on a Lebedev grid to an accuracy of $10^{-5}$ for all methods. This grid uses 194 angular points and radial grids containing either 50 or 200 points. One consistency check used to validate our software follows $\braket{r^0}$ to verify that it provides the correct number of unpaired electrons. Similarly, we verified that the analogous moment integral written in terms of the total density yields the total number of electrons. For several examples, we further verified that the numerically computed spin-density coincides with HFCC results for radial points near the origin.

The relative percentage error of predicted moments (average value) from several DFT functionals with respect to CCSD are tabulated, being defined as $\frac{|| X_{CCSD}| - |X_{DFT}|| }{ | X_{CCSD} |} * 100$. The alpha/beta and combined spin-densities, as well as the radial distribution function of the various moments as a function of radial distance, $r$, are provided. Numerical integration of the moment radial distribution functions yields the expectation value of the corresponding moment. Due to basis set-related convergence issues experienced in either software and the limited availability of specialized basis sets designed to capture HFCCs,  our analysis of the moment integral is partitioned into two test sets. Test set 1 incorporates second- and third-row elements and includes the B, C, N, O, F, Al, Si, P, S, and Cl atoms. Test set 2 is composed of the remaining atoms on the first two periods and includes the transition metals of the fourth row, that is, the Sc, Ti, V, Cr, Mn, Fe, Co, Ni, Cu, Li, Na, and K atoms. Atoms in Test set 1 use the aug-cc-pVTZ-J basis set,\cite{provasi2001effect,provasi2010optimized} whereas the atoms in Test set 2 use Roos' ANO double zeta basis set.\cite{pou1995density,widmark1990malmqvist,widmark1991density} Finally, we investigate the performance of the XC functionals in predicting anisotropic HFCCs. For this purpose, we employed the IGLO-III basis set.\cite{iglo3} The molecules considered in this step were \ce{AlO}, \ce{BO}, \ce{CH3}, \ce{CN }, \ce{CO+}, \ce{NH}, \ce{NH2}, and \ce{O2}. The aug-pc-4\cite{pc4} basis set is also used when appropriate. All basis sets were considered in their Cartesian forms.

To converge the SCF equations for DFT, the number of radial points in the integration grid was determined by ensuring that the radial integration accuracy was at least $10^{-6}$ ($\sim 10^{-5.66}$ for the transition metal atoms using the Roos double zeta basis set). In this context, every DFT calculation uses a Lebedev grid containing 770 angular points. Convergence at the SCF level was determined by ensuring that the norm of the energy and gradient fell below $10^{-9}$ and $10^{-8}$ a.u., respectively. Several KS-DFT functionals are studied, which are largely chosen based on their performance in previous work on HFCCs.\cite{windom2022benchmarking} They are listed in Table{~}\ref{tab:xcfunc}.

\begin{table}[t!]
\centering
\begin{tabular}{l l c c} 
        \toprule
     XC         &  Type            & $E_{X}^{HF} \%$ & Reference \\ \midrule
     HF         &  ---	          & 100     &   \citenum{hartree1928,fock1930,slater1930note}    \\
     SVWN5      &  LDA	          & 0	    &   \citenum{Slater1972,vwn} \\
     BLYP       &  GGA	          & 0	    &   \citenum{becke1988,lee1988}  \\
     OLYP       &  GGA	          & 0	    &   \citenum{handy2001left,lee1988}      \\
     XLYP       &  GGA	          & 0	    &   \citenum{Xin2004,lee1988}     \\
     PBE        &  GGA	          & 0       &   \citenum{perdew1996generalized}      \\
     M06-L      &  meta-GGA        & 0	    &   \citenum{Zhao2006}   \\
     B3LYP      &  Global-hybrid   & 20      &   \citenum{becke1988,becke1993,lee1988,Stephens1994}   \\
     O3LYP      &  Global-hybrid   & 11.61   &   \citenum{Cohen2021,lee1988}   \\  
     X3LYP      &  Global-hybrid   & 21.8    &   \citenum{Xin2004,lee1988}     \\
     PBE0       &  Global-hybrid   & 25      &   \citenum{Adamo1999}    \\  
     M06-2X     &  Global-hybrid   & 54	    &   \citenum{Zhao2008}   \\
     TPSSh      &  Global-hybrid   & 10      &   \citenum{staroverov2003}    \\  
     TPSS0      &  Global-hybrid   & 25	    &   \citenum{grimme2005accurate}     \\
     CAM-B3LYP  &  RSH             & 19-65   &   \citenum{yanai2004new}     \\
     HSE-03     &  RSH             & 25-0    &   \citenum{heyd2003hybrid} \\  
     HSE-06     &  RSH             & 25-0    &   \citenum{krukau2006influence}    \\
     LC-BLYP    &  RSH             & 0-100   &   \citenum{iikura2001long}   \\  
     QTP00      &  RSH	          & 54-91   &   \citenum{verma2014increasing}   \\
     QTP01      &  RSH	          & 23-100  &   \citenum{jin2016qtp}  \\
     QTP02      &  RSH	          & 28-100  &   \citenum{haiduke2018qtp2}    \\
     LC-QTP     &  RSH	          & 0-100	&   \citenum{haiduke2018qtp2}   \\
		\bottomrule
	\end{tabular}
 \caption{Exchange-correlation functionals used in this work, the exact exchange percentage ($E_{X}^{HF}$), and the respective references.}
 \label{tab:xcfunc}
\end{table}
\section{Results}

\subsection{A. Quality of the reference method}

An interesting aspect of isotropic hyperfine coupling constants is their dependence solely on the quality of the wavefunction (density/spin-density in the case of KS theory) at the nucleus. In this context, the quality of the reference wavefunction is assessed through the results presented in Table{~}\ref{tab:hfcc-wft}, which compares experimental HFCC values with those obtained from various wavefunction theory (WFT) methods, along with their respective average percent errors. The test set includes both neutral atoms and cations.

All the correlated methods studied in Table{~}\ref{tab:hfcc-wft} use UHF as the reference. So, relative to experimental values, the second-order perturbation reduces the percent error (compared to UHF) from 60.33 \% to 25.34 \%. However, although some improvement is observed in third-order, with an error dropping to 17.98 \%, moving from MBPT3 to MBPT4 does not significantly enhance the accuracy of the HFCC results (16.60 \%). On the other hand, Coupled-Cluster Theory with single and double excitations achieves the second best error estimate in comparison with the experiment, 11.08 \%. As expected, the most accurate result compared to the experimental HFCCs is obtained by also taking into account a perturbative correction for triple excitations, CCSD(T), with an error of 6.46 \%. Hence, compared to CCSD(T) data, CCSD deviates only by 5.48 \%, representing the best performance compared to alternative WFT methods such as MBPT4. Thus, considering the good performance of CCSD in predicting the HFCC results, we can proceed to the comparative evaluation of XC functionals using this method as a reference.

\begin{table*}[t!]
\begin{minipage}{\textwidth}
\centering
      \begin{adjustbox}{width=\textwidth}
\begin{tabular}{l S[table-format=-5.2] S[table-format=-5.2] S[table-format=-5.2] S[table-format=-5.2] S[table-format=-5.2] S[table-format=-5.2] S[table-format=-5.2]} 
        \toprule
               & {Exp.}   & {UHF}    &  {MBPT2}     & {MBPT3}   & {MBPT4}  & {CCSD}    & {CCSD(T)} \\ \midrule
\ce{B}         & 7.40     & 29.01    & -0.96        & -4.78     & -1.80     & 6.66      & 7.57      \\
\ce{C}         & 22.50    & 46.47    &  12.39       & 9.19      & 11.82     & 16.65     & 17.35     \\
\ce{N}         & 10.63    & 20.76    &  8.37        & 7.78      & 8.61      & 9.36      & 9.58      \\
\ce{O}         & 34.50    & -58.77   &  -27.43      & -27.17    & -28.59    & -29.95    & -30.29    \\
\ce{F}         & 301.70   & 544.20   &  276.21      & 282.57    & 287.50    & 295.70    & 297.28    \\
\ce{P}         & 61.00    & -81.85   &  -1.80       & 33.46     & 53.60     & 31.83     & 57.14     \\
\ce{Li}        & 401.80   & 399.68   &  409.03      & 412.03    & 413.15    & 413.47    & 413.71    \\
\ce{Na}        & 886.00   & 732.39   &  808.13      & 811.73    & 817.63    & 815.45    & 818.82    \\
\ce{Be+}       & -625.01  & -627.08  &  -634.22     & -636.20   & -636.76   & -636.88   & -636.95   \\
\ce{Mg+}       & -596.25  & -525.29  &  -551.07     & -552.45   & -553.35   & -552.83   & -553.39   \\
\ce{F}$^{6+}$  & 88890.00 & 87368.06 &  87506.58    & 87527.72  & 87530.51  & 87530.70  & 87530.90  \\
\ce{Mn+}       & 757.80   & 656.53   &  785.25      & 758.05    & 772.05    & 761.12    & 767.93    \\ \midrule
\%Error vs Exp. &         & 60.33    & 25.34        & 17.98     & 16.60     & 11.08     & 6.46 \\ \midrule
\%Error vs CCSD(T) &      & 68.62    & 20.63        & 13.52     & 11.19     & 5.48      &  \\\bottomrule
	\end{tabular}
       \end{adjustbox}
\caption{HFCC results\protect\footnote{The basis set used for B-P and F$^{6+}$ was aug-cc-pVTZ-J, while the basis set chosen for Li, Na, Be$^{+}$, Mg$^{+}$, Mn$^{+}$ was aug-pc-4. We retrieved experimental numbers from References{~}\citenum{kusch1949g,luther1949hyperfine,pendlebury1964hyperfine,harvey1965hyperfine,holloway1958determination,harvey1972diagonal,randolph1975measurement,ertmer1976zero,arimondo1977experimental,wineland1983laser,bergstrom1988radiative,pickering1996measurements,malkin2004scalar,bahramy2006first,kaupp2010hyperfine}.} of several wavefunction methods obtained for selected neutral atoms and cations, compared to experimental values (\protect\si{\mega\hertz}).}
 \label{tab:hfcc-wft} 
\end{minipage}
\end{table*}

\subsection{B. Test set 1}
\begin{figure*}[t!]
\includegraphics[width=0.90\linewidth]{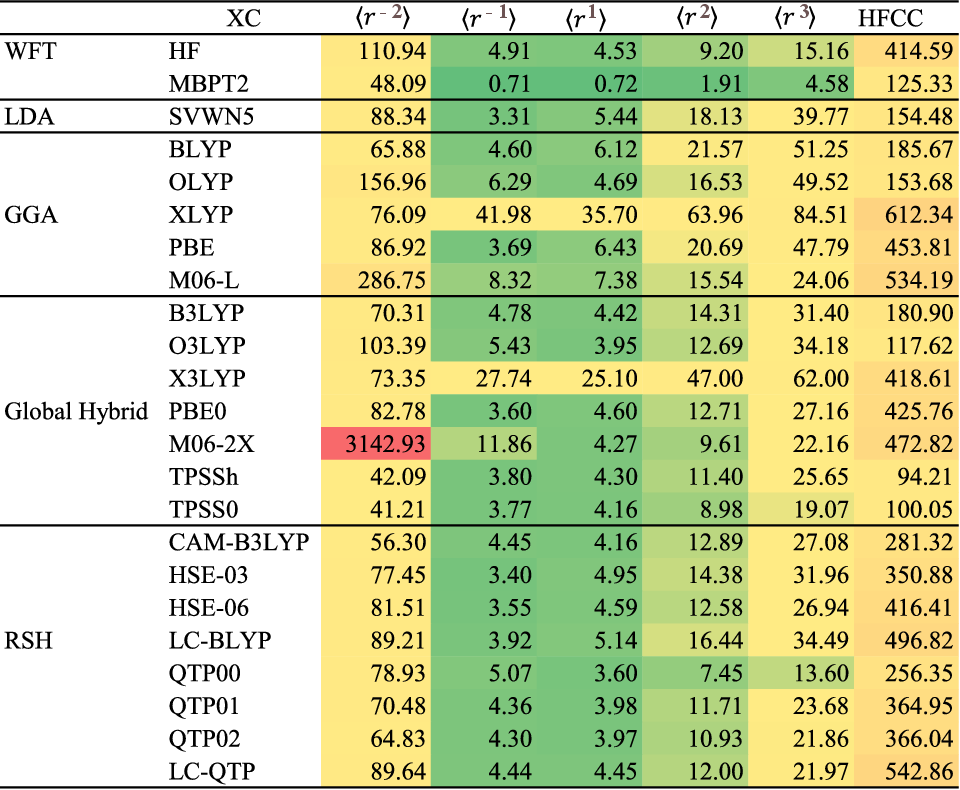}
\caption{Average percent error with respect to CCSD for each moment operator from spin density. The basis set used was aug-cc-pVTZ-J, and the atoms considered were B, C, N, O, F, Al, Si, P, S, and Cl.}
\label{fig:APE}
\end{figure*}

The average percentage errors provided by various DFT functionals are reported in Figure{~}\ref{fig:APE}. Considering the comparison with the full CCSD reference data, it becomes immediately clear that the calculated HFCCs from the majority of functionals and wavefunction methods (HF and MBPT2) are in error by over 100\% with respect to CCSD values, with the ``best" functionals being the MHGs of the TPSS-family (TPSSh and TPSS0) and O3LYP (HYB). Results from these functionals - all of which include some percentage of exact exchange - are rivaled by the presumably less-accurate SVWN5 (LDA) and OLYP (GGA) functionals for the atoms in this test set. Among RSHs, QTP00 seems more successful than the others for such HFCCs. The overall low quality of these results suggests an underlying deficiency in predicted spin-densities near the nucleus.

To verify this, we observe from the tabulated values of the short-range $\braket{r^{-2}}$ moment in Figure{~}\ref{fig:APE} that the errors across the board remain quite significant, although the aforementioned functionals of the TPSS family show the smallest deviations (41 - 42 \%). Again, CAM-B3LYP (56 \%), BLYP (66 \%) and B3LYP (70 \%) are quite successful, along with QTP02 and QTP01, which exhibit the next smallest errors amongst the RSH functionals (65 - 70 \%). We deduce from this that, on average, the aforementioned functionals produce spin-densities that are more locally accurate in regions that immediately surround the nucleus than the remaining DFAs considered and the worst error values are obtained from M06-L (287 \%) and M06-2X (3143 \%). Although MBPT2 is not so successful for HFCC predictions in this test set (error of 125 \%), the values for the $\braket{r^{-2}}$ moment are much better, with an average error of only 48 \%. Further analyzing this point, different trends of functional performance are loosely established  by the $\braket{r^{-1}}$ moment, although the  magnitude of errors found across all methods is significantly smaller,  making conclusive comparison more difficult. On average, M06-2X, X3LYP and XLYP are the clearly the worst DFAs for the $\braket{r^{-1}}$ moment (deviations between 12 and 42 \%). MBPT2 is quite accurate for this moment, with an error below 1\%. These seemingly paradoxical results can be understood by recognizing the following dichotomy regarding the KS-DFT spin-density: the accuracy of a method in predicting spin-densities at one point - the nucleus - does not necessarily imply that the spin-density is locally accurate in the immediate vicinity of the nucleus and vice versa.

\begin{figure}
\includegraphics[scale=0.65]{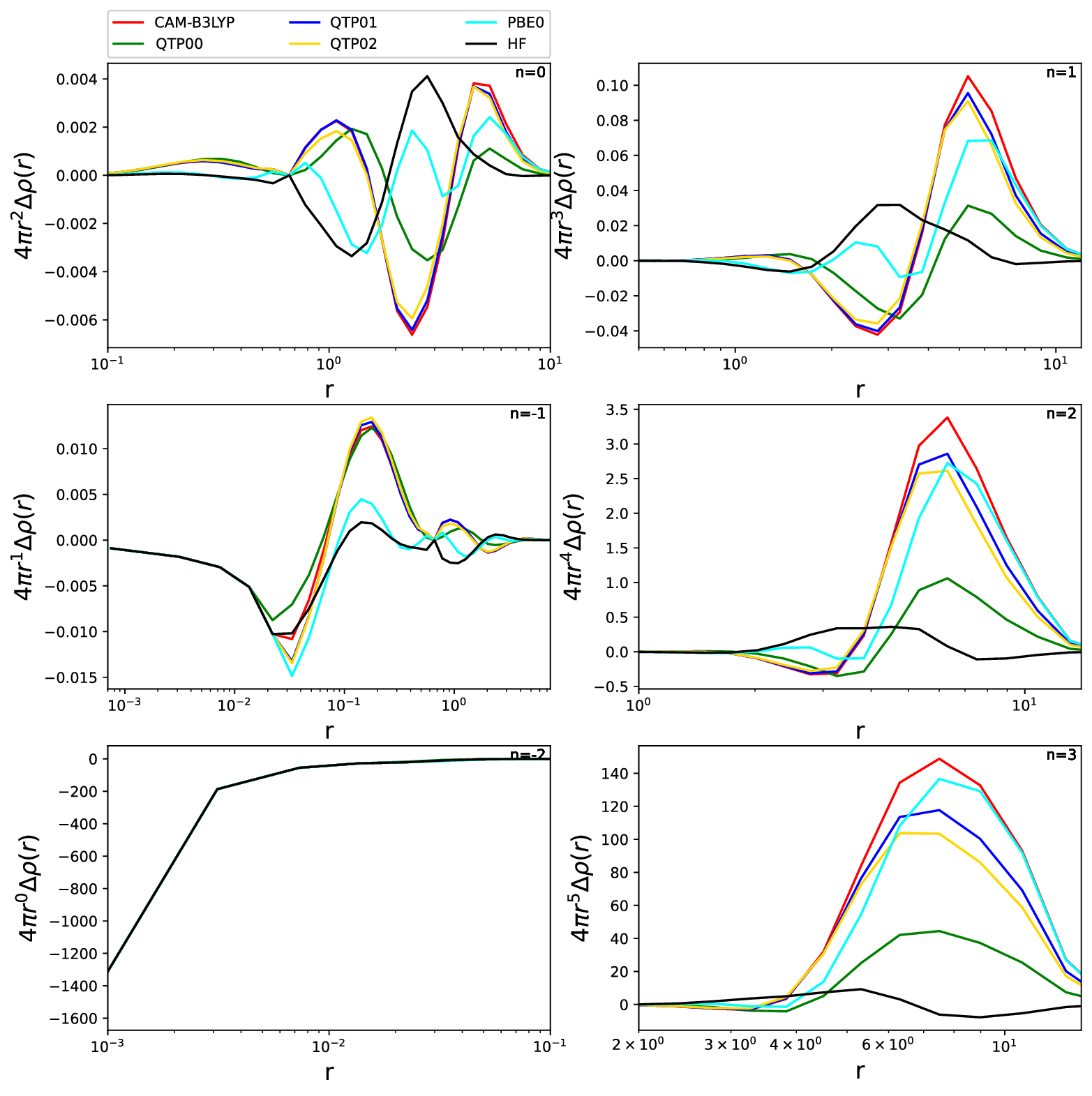}
\caption{Radial distribution of $\braket{r^n}$ for the Phosphorous atom. The y-axes are defined in terms of the total spin-density, which is assigned the shorthand variable $\rho (r)$ to reduce space. Every subplot has a y-axis that quantifies each method's deviation of the generalized expectation value against CCSD, seen to be $4\pi r^{2+n}\Delta \rho (r) = 4\pi r^{2+n} \rho^{CCSD} (r) -4\pi r^{2+n} \rho^{method} (r) $ for $n=-2, -1, 0, 1, 2, 3$. The radial distance, $r$, is measured in Bohr. }
\label{fig:PmomentsRS}
\end{figure}

To conceptualize these broad conclusions, Figure \ref{fig:PmomentsRS} quantifies the error in the radial distribution of the generalized expectation value, $\braket{r^n}$, in an effort to better understand spin-density deficiencies of the P atom as determined by the PBE0, CAM-B3LYP, QTP-DFT functionals, and HF theory. Error in the radial distribution function of the $r=0$ moment suggests that only HF and PBE0 generally agree closely with CCSD in short ranges surrounding the P nucleus, with the remaining functionals not too far off. However, in the intermediate and long-range regions, every method begins to deviate noticeably from CCSD, with the smallest deviation occurring for HF, as well as for the QTP00 and PBE0 functionals. For a shorter-range moment, $\braket{r^{-1}}$, broad agreement amongst the methods is forecast until radial distances around 0.005 Bohr and beyond, wherein PBE0 and HF seem to offer the overall closest agreement with CCSD.  For the shortest-range moment investigated, $\braket{r^{-2}}$, individual errors are superimposed and indiscernible, but one can notice that these errors are significant with respect to CCSD ($>$ 1000 \%). For long-range moments, $n > 0$, HF is clearly in best agreement overall, followed by QTP00. In this regime, it is also clear that CAM-B3LYP shows the worst overall performance. 

In addition, further analyzing long-range moments collected in Figure \ref{fig:APE}, $\braket{r^{1}}$,  $\braket{r^{2}}$, and  $\braket{r^{3}}$, it is clear that the QTP00 functional offers the best overall performance amongst the various DFAs studied, followed by the TPSS0. In this spatial regime, HF, M06-2X, TPSSh, PBE0, CAM-B3LYP, HSE-06, and the remaining QTP-like functionals (QTP01, QTP02, and LC-QTP) also show excellent agreement. In particular, the success of the M06-based functionals in the long-range is particularly noteworthy when considering the disparity in their short-range results, which are notably deplorable. Nevertheless, in general, the LDA (SVWN5), the collection of GGA functionals, and X3LYP exhibit the worst performance for such moments. The abysmal results of the LDA and GGA functionals are contrasted by the exceedingly successful results of HF for long-range spin-density moments, which even rival the quality of those of TPSS0. A quite impressive performance is noticed in MBPT2 results for these moments, which are the best ones in Figure \ref{fig:APE}.

\begin{figure}
\includegraphics[scale=0.65]{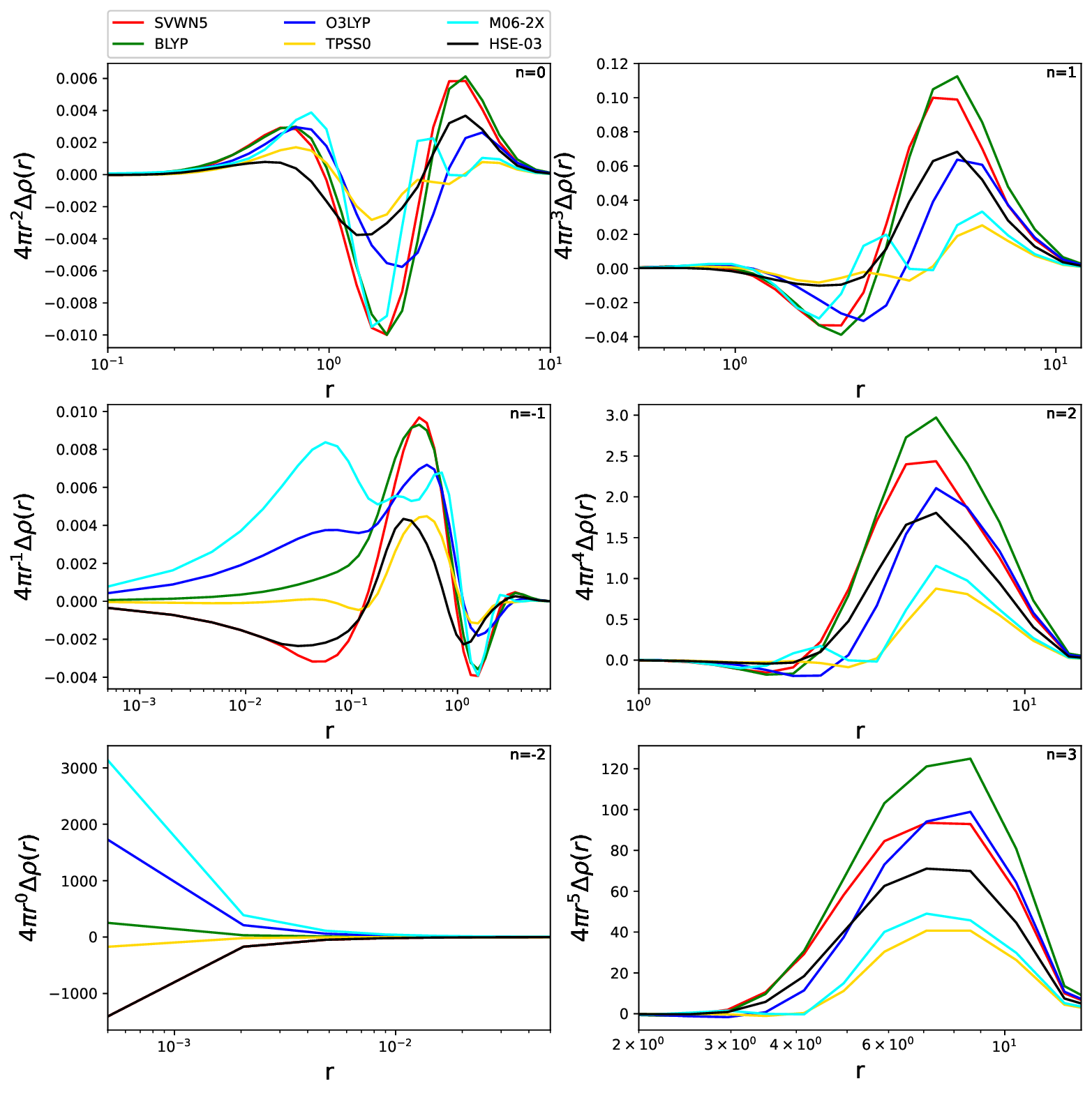}
\caption{Radial distribution of $\braket{r^n}$ for the Carbon atom. The y-axes are defined in terms of the total spin-density, which is assigned the short-hand variable $\rho (r)$ to reduce space. Every subplot has a y-axis that quantifies each method's deviation of the generalized expectation value against CCSD, seen to be $4\pi r^{2+n}\Delta \rho (r) = 4\pi r^{2+n} \rho^{CCSD} (r) -4\pi r^{2+n} \rho^{method} (r) $ for $n=-2, -1, 0, 1, 2, 3$. The radial distance, $r$, is measured in Bohr.}
\label{fig:CatomresultsOTHERS}
\end{figure}

Showing the error in the radial distribution of $\braket{r^n}$ for the C atom, Figure \ref{fig:CatomresultsOTHERS} reiterates the impressive performance of TPSS0 across the radial interval. We also include the HSE-03 RSH functional for posterity, as it seems to be a reliably consistent functional. It can be seen that, although the errors of M06-2X are quite large throughout the $r<0$ moments, the opposite is true for long-range moments, $r>0$. Furthermore, the overall error of SVWN5 is at least similar, if not slightly better, than the one from BLYP. 

\subsection{C. Test set 2}
\begin{figure*}[t!]
\includegraphics[width=0.90\linewidth]{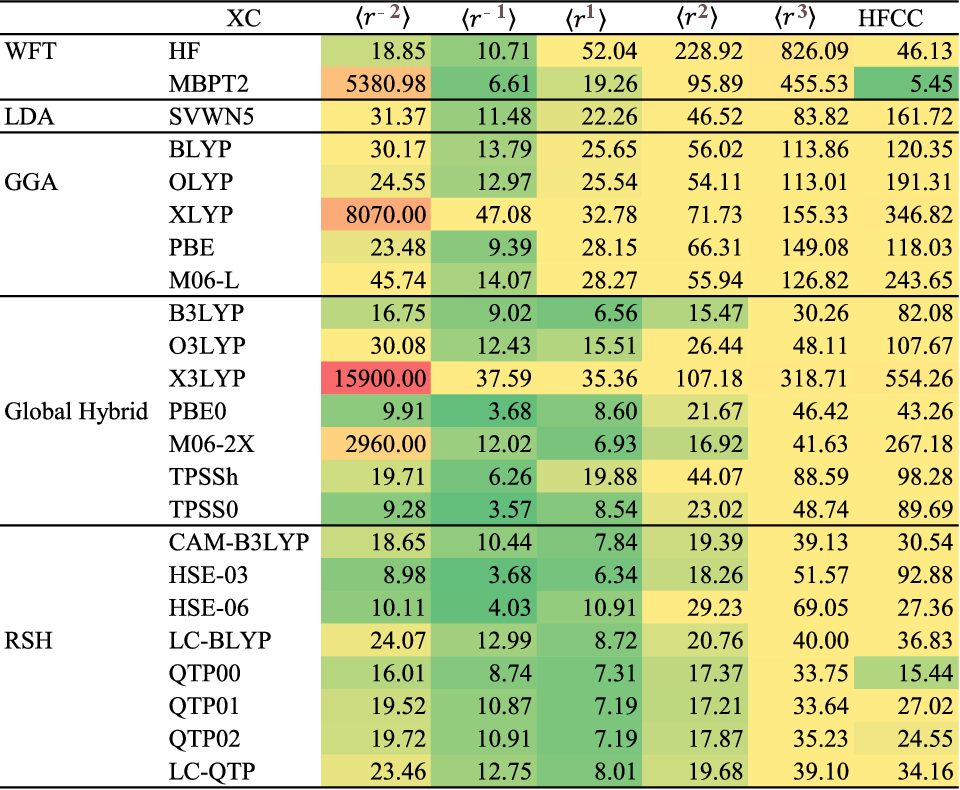}
\caption{Average percent error with respect to CCSD for each moment operator from spin density. The basis set used was the Roos Augmented Double Zeta ANO and the atoms considered were Sc, Ti, V, Cr, Mn, Fe, Co, Ni, Cu, Li, Na, and K.}
\label{fig:TMerrors}
\end{figure*}

A comparison of the average percentage error across a test set containing transition metal atoms is shown in Figure{~}\ref{fig:TMerrors}. Among the best-performing functionals, some noticeable trends in short-range moments, $\braket{r^{-2}}$ and $\braket{r^{-1}}$, stand out. As also seen in test set 1, all methods show better agreement with CCSD for the $\braket{r^{-1}}$ moment as compared to the $\braket{r^{-2}}$ moment, emphasizing each method's sensitivity to slight fluctuations in the spin-density that immediately surrounds the nucleus. In particular, the OLYP GGA functional shows respectable $\braket{r^{-2}}$ and $\braket{r^{-1}}$ results that broadly corroborate previous work noticing its excellent cost-to-accuracy ratio for larger organic radicals.\cite{windom2022benchmarking} The error in these two moments tends to improve by moving from GGAs to HF. Further improvements can be observed by moving from HF to B3LYP and PBE0 (HYBs) and next to TPSS0 (MHG). In particular, the PBE0 functional shows excellent agreement, which is in accordance with earlier work reporting its success in predicting HFCCs.\cite{windom2022benchmarking} The same is true for the TPSSh functional (MHG), wherein the error discrepancy with the analogous TPSS0 functional reinforces the delicate balance in adding HF exchange to MHG functionals. Overall, HSE-06, HSE-03, PBE0, and TPSS0 show comparable results.

\begin{figure}
\includegraphics[scale=0.65]{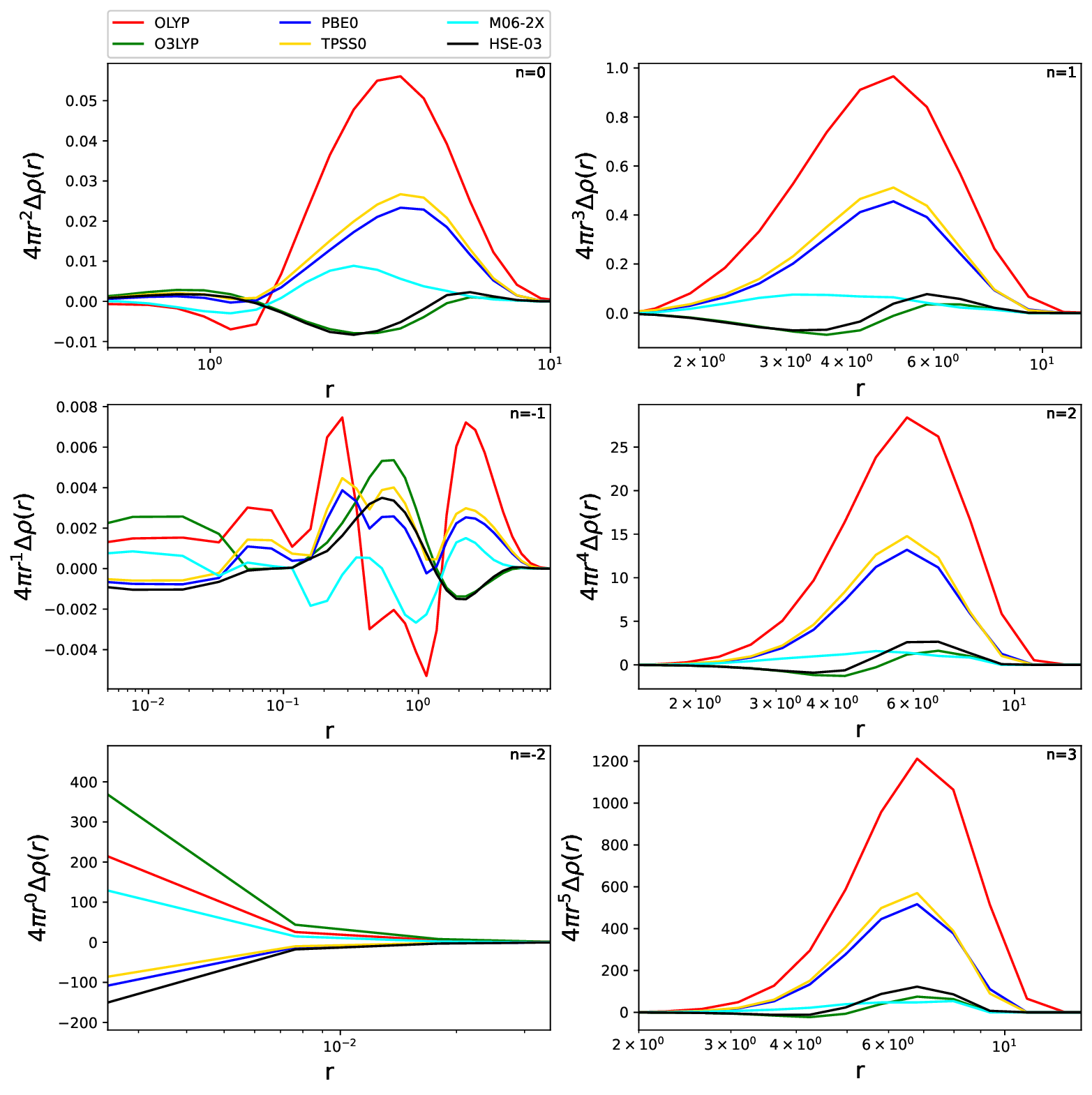}
\caption{Radial distribution of $\braket{r^n}$ for the Scandium atom. The y-axes are defined in terms of the total spin-density, which is assigned the short-hand variable $\rho (r)$ to reduce space. Every subplot has a y-axis that quantifies each method's deviation of the generalized expectation value against CCSD, seen to be $4\pi r^{2+n}\Delta \rho (r) = 4\pi r^{2+n} \rho^{CCSD} (r) -4\pi r^{2+n} \rho^{method} (r) $ for $n=-2, -1, 0, 1, 2, 3$. The radial distance, $r$, is measured in Bohr.}
\label{fig:ScatomOTHER}
\end{figure}

\begin{figure}
\includegraphics[scale=0.65]{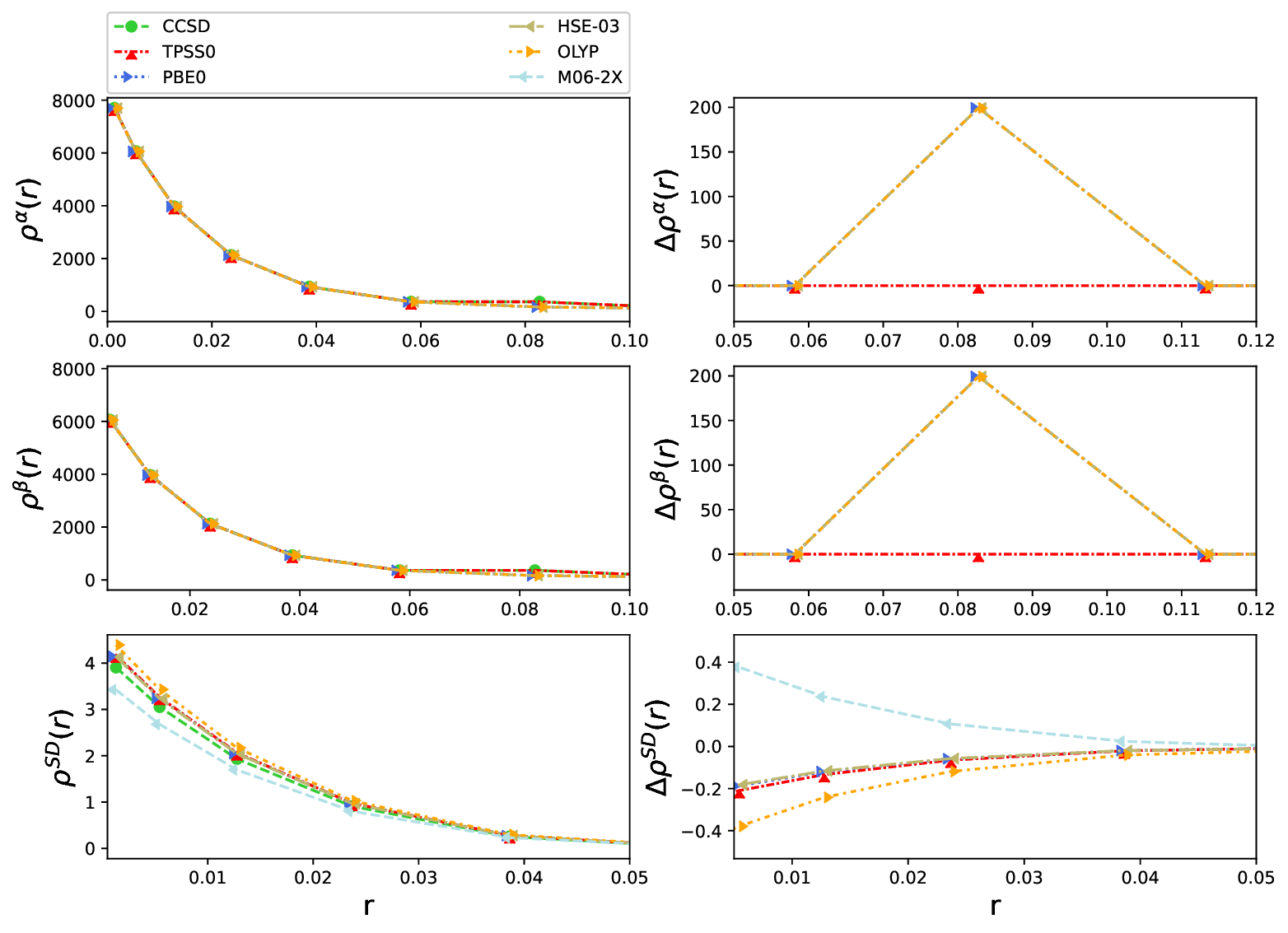}
\caption{Analysis of the Cu atom's global spin-density, $\rho^{SD}(r)$, as predicted by select functionals and CCSD. The first column of plots further records the alpha and beta contributions to the spin-density, denoted by y-axis labels $\rho^{\alpha}(r)$ and $\rho^{\beta}(r)$ respectively.  The second column of plots quantifies each functionals' error as compared to CCSD for the alpha ($\Delta \rho^{\alpha}(r)$) and beta ($\Delta \rho^{\beta}(r)$) electron densities, as well as the global spin-density error ($\Delta \rho^{SD}(r)$). The radial distance, $r$, is measured in Bohr. Note: M06-2X is only shown in the total spin-density and its error plots.}
\label{fig:CuatomALL}
\end{figure}

\begin{figure}
\includegraphics[scale=0.65]{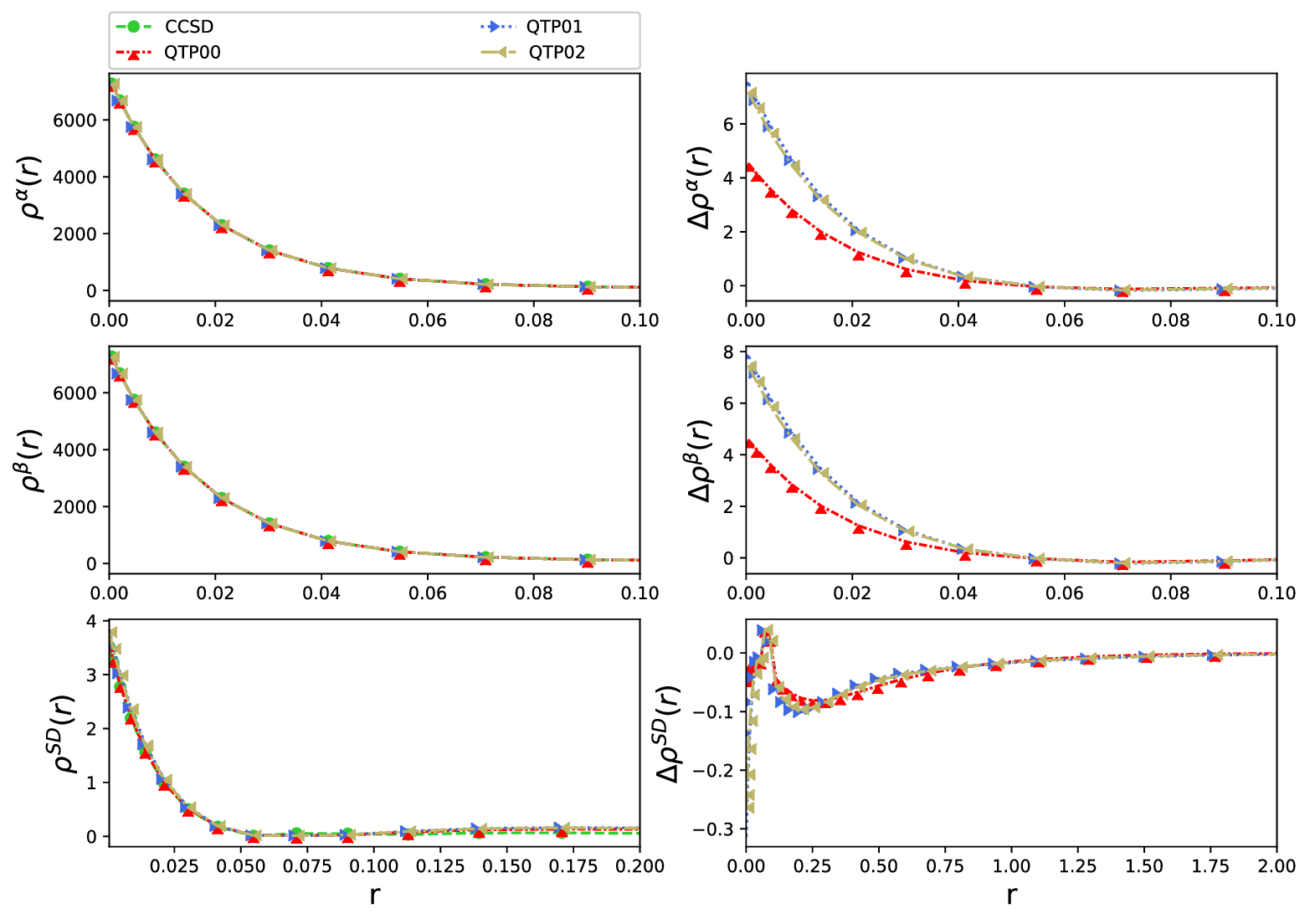}
\caption{Analysis of the Ni atom's global spin-density, $\rho^{SD}(r)$, as predicted by select RSH functionals and CCSD. The first column of plots further records the alpha and beta contributions to the spin-density, denoted by y-axis labels $\rho^{\alpha}(r)$ and $\rho^{\beta}(r)$ respectively.  The second column of plots quantifies the error RSH functionals show as compared to CCSD for the alpha ($\Delta \rho^{\alpha}(r)$) and beta ($\Delta \rho^{\beta}(r)$) electron densities, as well as the global spin-density error ($\Delta \rho^{SD}(r)$). The radial distance, $r$, is measured in Bohr. }
\label{fig:NiatomRSH}
\end{figure}

 In order to deeply investigate some of these functionals, Figure \ref{fig:ScatomOTHER} reports the error in the radial distribution of the various moments for the Sc atom. From this particular example, it is immediately clear that the overall worst-performing functional is OLYP, emphasizing the lack of a clear relationship between HFCC predictions and global spin-density performance. On the other hand, once exact exchange is added to form the O3LYP functional, dramatic improvements in the global spin-density are recovered, arguably providing better quality than PBE0, except for the shortest-range moment (n = -2). From this example, we also notice that the notoriously poor-performing M06-2X exhibits significant improvements in both short- and long-range spin-densities and even competes with HSE-03 as being the functional having a better accordance with CCSD results. 

Figure \ref{fig:CuatomALL} lends further perspective into these findings by quantifying the spin-densities and their errors for the Cu atom. In this case, the spin-densities are generally overestimated as compared to CCSD, with M06-2X being the only exception. Nevertheless, M06-2X exhibits the worst deviations amongst those shown, which is also verified by the extreme error magnitudes shown in Figure \ref{fig:TMerrors} for this DFA (particularly for $r^{-2}$). Furthermore, the $\alpha$/$\beta$ spin-density error with respect to CCSD  appreciably increases only around 0.08 Bohr and TPSS0 is a relevant exception to this trend. Anyway, the error in these individual $\alpha$/$\beta$ spin-densities largely cancels along this interval once combined to form the global spin-density. In fact, the total spin-density error with respect to CCSD is effectively nonexistent starting slightly before 0.05 Bohr.

Comparing the errors for RSHs, the HSE-03 and HSE-06 functionals show the overall best performance for short-range moments. Conversely, the QTP-family of functionals exhibits the smallest errors in the long-range of $n>0$ moments. This highlights the differences in design philosophy between QTP and the HSE functionals. The latter includes a percentage of exact exchange at small $r$, but GGA exchange otherwise, whereas the QTP-type of functionals includes exact exchange at large $r$ and GGA exchange otherwise. Clearly, there are trade-offs between these philosophies for transition metals, although it should also be pointed out that QTP functionals offer better agreement in predicted HFCCs. Despite earlier work showing excellent overall performance of QTP01 and QTP02  functionals for HFCCs of transition metal complexes with respect to QTP00, the current study now evidences the QTP00 functional as being in better agreement with CCSD for individual atoms as compared to the other members of the QTP family. 

In order to investigate this contradiction, Figure \ref{fig:NiatomRSH} depicts the $\alpha$, $\beta$, and total spin-densities as well as the errors for each of them. From this, we notice that QTP00 predicts $\alpha$/$\beta$ spin-densities that are overwhelmingly in closer agreement with CCSD than the others. Because spin-densities have units of $a.u.^{-3}$, a small spin-density discrepancy with exact results can be magnified in the corresponding property calculations. At this point, the fact that QTP00 shows the best overall agreement with CCSD for HFCCs of TM atoms, which are only in error by 15\%, seems to be directly related to the fact that it consistently yields competitive global spin-densities, particularly at the nucleus. This is in opposition to many functional trends discussed throughout this work.

It is also interesting to point out that HF offers a competitive agreement with CCSD for TM HFCCs and even surpasses every member of the LDA and GGA families to compete with HYB and MHG functionals. Furthermore, impressive agreement with CCSD is now observed for the $\braket{r^{-2}}$ and $\braket{r^{-1}}$ moments, in spite of HF lacking any formal accounting for electron correlation effects. On the other hand, HF is no longer a good choice for long-range moments (n $>$ 1) as it was for test set 1, now yielding the largest deviations in Figure \ref{fig:TMerrors}. Perhaps, these findings evidence a warning against several DFAs, exposing the fragility of developing exchange-correlation potentials that are devoid of the rigorous considerations stated by many-body theory and COT. Another point deserving some comments is the finding that MBPT2 is now inadequate both for the shortest-range moment, $\braket{r^{-2}}$, as for the longest-range one, $\braket{r^{3}}$. Curiously, MBPT2 still provides the best HFCC predictions, with an error of only 5\%. 

In addition, with respect to long-range moments, $\braket{r^{1}}$, $\braket{r^2}$, and $\braket{r^{3}}$,  the errors accumulated by M06-2X are significantly improved as compared to its performance in the short-range spin-density analysis. Otherwise, the aforementioned trends are largely held by OLYP, PBE0, and TPSS0 - all of which continue to be among the best performers in their respective functional families. In this region, QTP00, QTP01, and QTP02 show the best overall agreement of the RSH family, and such a decision arises from the significant increase in the rate at which the HSE series of functionals accumulates error upon moving to the $\braket{r^{3}}$ moment.

\subsection{D. Anisotropic HFCC}

\begin{figure*}
    \centering
    \includegraphics[width=0.6\linewidth]{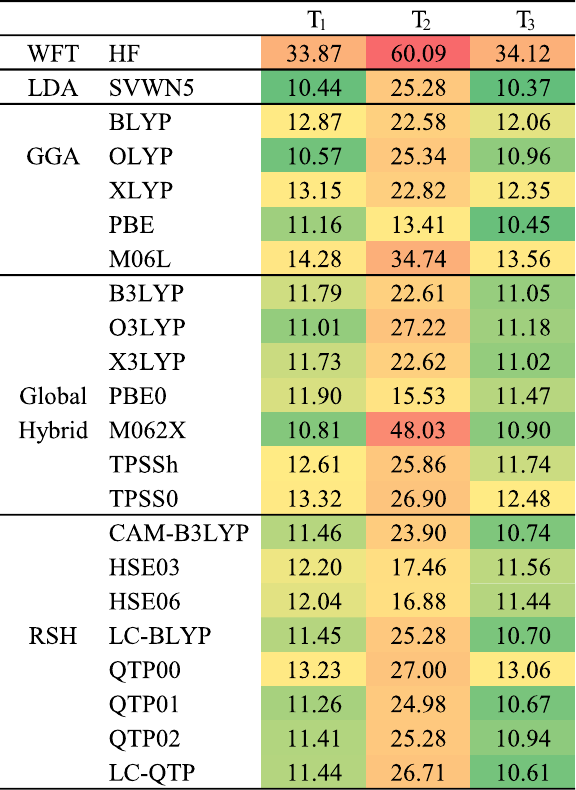}
    \caption{Percent errors (\%E) with respect to the experimental anisotropic (T$_1$-T$_3$) contributions to the hyperfine coupling constants of small molecules. The IGLO-III basis set was used,\cite{iglo3} as prescribed in Ref.{~}\citenum{aniso-bench}.}
    \label{fig:fermi-aniso}
\end{figure*}

Finally, Figure~\ref{fig:fermi-aniso} shows the percent errors for the anisotropic spin-dipolar components (T$_1$–T$_3$) of the HFCCs. Experimental geometries, retrieved from the NIST,\cite{Russell_2018} were used for all systems. The calculated values are compared with the corresponding experimental data.\cite{aniso-bench} The complete numerical results are provided in the Supplementary Material. Overall, the heat map shows that the calculated T$_1$ and T$_3$ components are in better agreement with experiment than T$_2$. For example, excluding HF, all functionals give percent errors between \SIrange{10.44}{14.28}{\percent} for T$_1$ and between \SIrange{10.37}{13.56}{\percent} for T$_3$. In contrast, the errors for T$_2$ are generally larger, with the best result obtained using the pure GGA functional PBE, which gives a percent error of \SI{13.41}{\percent}. The QTP family of functionals ranks among the best performers for T$_1$ and T$_3$, although its performance for T$_2$ is only average.

\section{Conclusion}

In this work, we have reported the deviations of the HF method and various DFT functionals in predicting the HFCCs and $r=-2, -1, 1, 2, 3$ moments for atoms in the second, third, and fourth row as compared to CCSD. Although many of these functionals accurately predict HFCCs, thereby implying that the corresponding functional exhibits an accurate spin-density at the atomic nucleus, we emphasize that general conclusions about the resulting global spin-density cannot be inferred based exclusively upon this particular point in space. By inspecting error plots of each moment's radial distribution function, we have shown that common DFT functionals largely fail to reproduce CCSD-quality global spin-densities, particularly within a range of 1 Bohr of the nucleus. This is corroborated by analyzing errors in the $\alpha$/$\beta$ and combined spin-densities as a function of radial distance.  It is known that dramatic fluctuations in the (spin) density are likely to occur in regions immediately surrounding the core and, in such a context, a method's ability to accurately account for electron exchange-correlation effects becomes significantly more important. Consequently, improvements on current DFT approximations demand an accurate KS potential, $V_s = \frac{\delta E}{\delta \rho(r)}$, from which the orbitals and, by extension, the global spin-densities, are constructed.

The governing design philosophy behind the QTP functionals is to construct a potential by the re-parameterization of existing functionals (like CAM-B3LYP) in order to obey the correlated orbital theory statements. In essence, following this protocol will certify that the resulting potential is accurate. The fact that QTP00 offers a consistent improvement over CAM-B3LYP  across most of the error metrics studied in this work, in addition to consistently outperforming the overwhelming majority of tested functionals, indicates that there are benefits from enforcing rigorous conditions on functional parameterization.  

Nevertheless, unanimous agreement in predicted HFCCs and spin-densities by any functional is not found as compared to CCSD. Although we can notice that there are benefits in improving the exchange-correlation potential by satisfying the IP theorem (QTP00 performs consistently well overall), it still shows significant deviations with respect to CCSD. Unfortunately, the functionals explored in this study lack a systematic route of improvement toward the FCI limit. The results of this work motivate further exploration of rigorous constraints that can be applied during parameterization.  In this context, a possible way to expand upon the existing QTP-DFT parameterization/design philosophy might be to also insist that the resulting functional approximation yields accurate spin-densities at the nucleus. 
Beyond minimizing the impact of issues associated with the Devils' Triangle of KS-DFT, a consequence of adding this constraint to the parameterization would at least yield a functional that should be well-suited for capturing the HFCCs. By adding this requirement to the list of QTP constraints on functional parameterization, marginal improvements in the short-range spin-density may also be recovered, and this will be the focal point of future work.

To briefly summarize the general trends found in the current work:

\begin{itemize}
    \item The errors in calculated HFCCs obtained by the GGA OLYP  for test set 1 are comparatively small, although the analysis of its deviations in both the short- and long-range moments reveals appreciable errors. Moreover, OLYP is not particularly accurate for HFCCs of test set 2, including transition metal elements, although this DFA now shows respectable $\braket{r^{-2}}$ and $\braket{r^{-1}}$ results. This implies that the overall quality of its global spin-density is generally poor, despite this functional normally provides accurate HFCC results for small/large organic molecules and transition metal complexes.\cite{windom2022benchmarking}
   
    \item The impact of adding some percentage of exact HF exchange is noticed in previous work\cite{windom2022benchmarking} and again reiterated in the current study. Perhaps the most illustrative examples of this point are the results of the GGA OLYP and its HYB counterpart, O3LYP. In the case of HFCCs, for example, the predicted values of O3LYP improve upon OLYP results roughly by $30\%$ in test set 1 and approximately by $80\%$ in test set 2. Furthermore, by comparing the errors across every moment in both test sets,  the superiority of the O3LYP spin-density over that from OLYP becomes evident. Similar trends are also generally observed for the BLYP/B3LYP and PBE/PBE0 pairs of GGA/HYB functionals.

     \item The accuracy of the TPSS-type of MHG functionals and of PBE0 (HYB) is again highlighted. Also,  each QTP-DFT  functional individually stands out in different radial regimes. However, we notice that the overall best functional in this group is QTP00. 
\end{itemize}

\section{Acknowledgments}

R. L. A. H. acknowledges S\~ao Paulo Research Foundation (FAPESP) for support (grant 2022/05138-0) and Brazilian National Council for Scientific and Technological Development (CNPq) for financial funding (grant 304407/2024-0). The QTP members (Z.W.W, A.P., R.J.B.) acknowledge support by the Air Force Office of Scientific Research under AFOSR Award No. FA9550—23-1-0118. Z. W. W. also thanks the National Science Foundation and the Molecular Sciences Software Institute for financial support under Grant No. CHE-2136142

\clearpage

\section*{References}

\bibliography{ref.bib}

@article{bartlett2019adventures,
  title={Adventures in DFT by a wavefunction theorist},
  author={Bartlett, Rodney J},
  journal={The Journal of chemical physics},
  volume={151},
  number={16},
  pages={160901},
  year={2019},
  publisher={AIP Publishing LLC}
}

@article{munzarova2000mechanisms,
  title={Mechanisms of EPR hyperfine coupling in transition metal complexes},
  author={Munzarov{\'a}, Mark{\'e}ta L and Kub{\'a}cek, Pavel and Kaupp, Martin},
  journal={Journal of the American Chemical Society},
  volume={122},
  number={48},
  pages={11900--11913},
  year={2000},
  publisher={ACS Publications}
}

@article{remenyi2007density,
  title={Density functional study of EPR parameters and spin-density distribution of azurin and other blue copper proteins},
  author={Remenyi, Christian and Reviakine, Roman and Kaupp, Martin},
  journal={The Journal of Physical Chemistry B},
  volume={111},
  number={28},
  pages={8290--8304},
  year={2007},
  publisher={ACS Publications}
}

@article{remenyi2006density,
  title={Density functional study of electron paramagnetic resonance parameters and spin density distributions of dicopper (I) complexes with bridging azo and tetrazine radical-anion ligands},
  author={Remenyi, Christian and Reviakine, Roman and Kaupp, Martin},
  journal={The Journal of Physical Chemistry A},
  volume={110},
  number={11},
  pages={4021--4033},
  year={2006},
  publisher={ACS Publications}
}

@article{hedegard2013validating,
  title={Validating and analyzing EPR hyperfine coupling constants with density functional theory},
  author={Hedeg{\aa}rd, Erik D and Kongsted, Jacob and Sauer, Stephan PA},
  journal={Journal of Chemical Theory and Computation},
  volume={9},
  number={5},
  pages={2380--2388},
  year={2013},
  publisher={ACS Publications}
}

@article{ranasinghe2017note,
  title={A note on the accuracy of KS-DFT densities},
  author={Ranasinghe, Duminda S and Perera, Ajith and Bartlett, Rodney J},
  journal={The Journal of Chemical Physics},
  volume={147},
  number={20},
  pages={204103},
  year={2017},
  publisher={AIP Publishing LLC}
}

@article{kohn1965self,
  title={Self-consistent equations including exchange and correlation effects},
  author={Kohn, Walter and Sham, Lu Jeu},
  journal={Physical review},
  volume={140},
  number={4A},
  pages={A1133},
  year={1965},
  publisher={APS}
}

@article{jacob2012spin,
  title={Spin in density-functional theory},
  author={Jacob, Christoph R and Reiher, Markus},
  journal={International Journal of Quantum Chemistry},
  volume={112},
  number={23},
  pages={3661--3684},
  year={2012},
  publisher={Wiley Online Library}
}

@article{reiher2007definition,
  title={On the definition of local spin in relativistic and nonrelativistic quantum chemistry},
  author={Reiher, Markus},
  journal={Faraday Discussions},
  volume={135},
  pages={97--124},
  year={2007},
  publisher={Royal Society of Chemistry}
}

@article{perera1994theoretical,
  title={A theoretical study of hyperfine coupling constants},
  author={Perera, S Ajith and Watts, John D and Bartlett, Rodney J},
  journal={The Journal of chemical physics},
  volume={100},
  number={2},
  pages={1425--1434},
  year={1994},
  publisher={American Institute of Physics}
}

@article{perera1996electron,
  title={Electron correlation effects on the theoretical calculation of nuclear magnetic resonance spin--spin coupling constants},
  author={Perera, S Ajith and Nooijen, Marcel and Bartlett, Rodney J},
  journal={The Journal of chemical physics},
  volume={104},
  number={9},
  pages={3290--3305},
  year={1996},
  publisher={American Institute of Physics}
}

@article{perera1994coupled,
  title={Coupled-cluster calculations of indirect nuclear coupling constants: The importance of non-Fermi contact contributions},
  author={Perera, S Ajith and Sekino, Hideo and Bartlett, Rodney J},
  journal={The Journal of chemical physics},
  volume={101},
  number={3},
  pages={2186--2191},
  year={1994},
  publisher={American Institute of Physics}
}

@article{verma2013massively,
  title={Massively parallel implementations of coupled-cluster methods for electron spin resonance spectra. I. Isotropic hyperfine coupling tensors in large radicals},
  author={Verma, Prakash and Perera, Ajith and Morales, Jorge A},
  journal={The Journal of chemical physics},
  volume={139},
  number={17},
  pages={174103},
  year={2013},
  publisher={American Institute of Physics}
}

@article{perera2017benchmark,
  title={Benchmark coupled-cluster g-tensor calculations with full inclusion of the two-particle spin-orbit contributions},
  author={Perera, Ajith and Gauss, J{\"u}rgen and Verma, Prakash and Morales, Jorge A},
  journal={The Journal of Chemical Physics},
  volume={146},
  number={16},
  pages={164104},
  year={2017},
  publisher={AIP Publishing LLC}
}

@article{rappoport2009functional,
  title={Which functional should I choose?},
  author={Rappoport, Dmitrij  // i\kern -.15em j and Crawford, Nathan RM and Furche, Filipp and Burke, Kieron and Wiley, C},
  journal={Computational Inorganic and Bioinorganic Chemistry, Wiley-Blackwell},
  year={2009}
}

@article{perdew1992accurate,
  title={Accurate and simple analytic representation of the electron-gas correlation energy},
  author={Perdew, John P and Wang, Yue},
  journal={Physical review B},
  volume={45},
  number={23},
  pages={13244},
  year={1992},
  publisher={APS}
}

@article{ceperley1980ground,
  title={Ground state of the electron gas by a stochastic method},
  author={Ceperley, David M and Alder, Berni J},
  journal={Physical review letters},
  volume={45},
  number={7},
  pages={566},
  year={1980},
  publisher={APS}
}

@article{becke1993new,
  title={A new mixing of Hartree--Fock and local density-functional theories},
  author={Becke, Axel D},
  journal={The Journal of chemical physics},
  volume={98},
  number={2},
  pages={1372--1377},
  year={1993},
  publisher={American Institute of Physics}
}

@article{perdew1996generalized,
  title={Generalized gradient approximation made simple},
  author={Perdew, John P and Burke, Kieron and Ernzerhof, Matthias},
  journal={Physical review letters},
  volume={77},
  number={18},
  pages={3865},
  year={1996},
  publisher={APS}
}

@article{tao2003climbing,
  title={Climbing the density functional ladder: Nonempirical meta--generalized gradient approximation designed for molecules and solids},
  author={Tao, Jianmin and Perdew, John P and Staroverov, Viktor N and Scuseria, Gustavo E},
  journal={Physical Review Letters},
  volume={91},
  number={14},
  pages={146401},
  year={2003},
  publisher={APS}
}

@article{bartlett2017power,
  title={The power of exact conditions in electronic structure theory},
  author={Bartlett, Rodney J and Ranasinghe, Duminda S},
  journal={Chemical Physics Letters},
  volume={669},
  pages={54--70},
  year={2017},
  publisher={Elsevier}
}

@article{windom2022examining,
  title={Examining fundamental and excitation gaps at the thermodynamic limit: A combined (QTP) DFT and coupled cluster study on trans-polyacetylene and polyacene},
  author={Windom, Zachary W and Perera, Ajith and Bartlett, Rodney J},
  journal={The Journal of Chemical Physics},
  volume={156},
  number={20},
  pages={204308},
  year={2022},
  publisher={AIP Publishing LLC}
}

@article{windom2022benchmarking,
  title={Benchmarking isotropic hyperfine coupling constants using (QTP) DFT functionals and coupled cluster theory},
  author={Windom, Zachary W and Perera, Ajith and Bartlett, Rodney J},
  journal={The Journal of Chemical Physics},
  volume={156},
  number={9},
  pages={094107},
  year={2022},
  publisher={AIP Publishing LLC}
}

@article{oliphant1994systematic,
  title={A systematic comparison of molecular properties obtained using Hartree--Fock, a hybrid Hartree--Fock density-functional-theory, and coupled-cluster methods},
  author={Oliphant, Nevin and Bartlett, Rodney J},
  journal={The Journal of chemical physics},
  volume={100},
  number={9},
  pages={6550--6561},
  year={1994},
  publisher={American Institute of Physics}
}

@article{bartlett2005ab,
  title={Ab initio density functional theory: The best of both worlds?},
  author={Bartlett, Rodney J and Lotrich, Victor F and Schweigert, Igor V},
  journal={The Journal of chemical physics},
  volume={123},
  number={6},
  pages={062205},
  year={2005},
  publisher={American Institute of Physics}
}

@article{talman1976optimized,
  title={Optimized effective atomic central potential},
  author={Talman, James D and Shadwick, William F},
  journal={Physical Review A},
  volume={14},
  number={1},
  pages={36},
  year={1976},
  publisher={APS}
}

@article{hirata2001can,
  title={Can optimized effective potentials be determined uniquely?},
  author={Hirata, So and Ivanov, Stanislav and Grabowski, Ireneusz and Bartlett, Rodney J and Burke, Kieron and Talman, James D},
  journal={The Journal of Chemical Physics},
  volume={115},
  number={4},
  pages={1635--1649},
  year={2001},
  publisher={American Institute of Physics}
}

@article{chong2002interpretation,
  title={Interpretation of the Kohn--Sham orbital energies as approximate vertical ionization potentials},
  author={Chong, Delano P and Gritsenko, Oleg V and Baerends, Evert J},
  journal={The Journal of Chemical Physics},
  volume={116},
  number={5},
  pages={1760--1772},
  year={2002},
  publisher={American Institute of Physics}
}

@article{mendes2021devil,
  title={The Devil’s Triangle of Kohn--Sham density functional theory and excited states},
  author={Mendes, Rodrigo A and Haiduke, Roberto L A and Bartlett, Rodney J},
  journal={The Journal of Chemical Physics},
  volume={154},
  number={7},
  pages={074106},
  year={2021},
  publisher={AIP Publishing LLC}
}

@article{mendes2021performance,
  title={Performance of new exchange--correlation functionals in providing vertical excitation energies of metal complexes},
  author={Mendes, Rodrigo Araujo and Haiduke, Roberto Luiz Andrade},
  journal={Theoretical Chemistry Accounts},
  volume={140},
  number={11},
  pages={146},
  year={2021},
  publisher={Springer},
  url = {https://doi.org/10.1007/s00214-021-02844-8},
  doi = {10.1007/s00214-021-02844-8}
}

@article{gudmundsdottir2013self,
  title={Self-interaction corrected density functional calculations of molecular Rydberg states},
  author={Gudmundsd{\'o}ttir, Hildur and Zhang, Yao and Weber, Peter M and J{\'o}nsson, Hannes},
  journal={The Journal of chemical physics},
  volume={139},
  number={19},
  pages={194102},
  year={2013},
  publisher={American Institute of Physics}
}

@article{baer2010tuned,
  title={Tuned range-separated hybrids in density functional theory},
  author={Baer, Roi and Livshits, Ester and Salzner, Ulrike},
  journal={Annual review of physical chemistry},
  volume={61},
  pages={85--109},
  year={2010}
}

@article{dreuw2003long,
  title={Long-range charge-transfer excited states in time-dependent density functional theory require non-local exchange},
  author={Dreuw, Andreas and Weisman, Jennifer L and Head-Gordon, Martin},
  journal={The Journal of chemical physics},
  volume={119},
  number={6},
  pages={2943--2946},
  year={2003},
  publisher={American Institute of Physics}
}

@article{mori2014derivative,
  title={The derivative discontinuity of the exchange--correlation functional},
  author={Mori-S{\'a}nchez, Paula and Cohen, Aron J},
  journal={Physical Chemistry Chemical Physics},
  volume={16},
  number={28},
  pages={14378--14387},
  year={2014},
  publisher={Royal Society of Chemistry}
}

@article{perdew1982density,
  title={Density-functional theory for fractional particle number: derivative discontinuities of the energy},
  author={Perdew, John P and Parr, Robert G and Levy, Mel and Balduz Jr, Jose L},
  journal={Physical Review Letters},
  volume={49},
  number={23},
  pages={1691},
  year={1982},
  publisher={APS}
}

@article{hirata2002time,
  title={Time-dependent density functional theory employing optimized effective potentials},
  author={Hirata, So and Ivanov, Stanislav and Grabowski, Ireneusz and Bartlett, Rodney J},
  journal={The Journal of chemical physics},
  volume={116},
  number={15},
  pages={6468--6481},
  year={2002},
  publisher={American Institute of Physics}
}

@article{bartlett2009towards,
  title={Towards an exact correlated orbital theory for electrons},
  author={Bartlett, Rodney J},
  journal={Chemical Physics Letters},
  volume={484},
  number={1-3},
  pages={1--9},
  year={2009},
  publisher={Elsevier}
}

@article{zhao2008m06,
  title={The M06 suite of density functionals for main group thermochemistry, thermochemical kinetics, noncovalent interactions, excited states, and transition elements: two new functionals and systematic testing of four M06-class functionals and 12 other functionals},
  author={Zhao, Yan and Truhlar, Donald G},
  journal={Theoretical chemistry accounts},
  volume={120},
  number={1},
  pages={215--241},
  year={2008},
  publisher={Springer}
}

@article{yanai2004new,
  title={A new hybrid exchange--correlation functional using the Coulomb-attenuating method (CAM-B3LYP)},
  author={Yanai, Takeshi and Tew, David P and Handy, Nicholas C},
  journal={Chemical physics letters},
  volume={393},
  number={1-3},
  pages={51--57},
  year={2004},
  publisher={Elsevier}
}

@article{jin2016qtp,
  title={The QTP family of consistent functionals and potentials in Kohn-Sham density functional theory},
  author={Jin, Yifan and Bartlett, Rodney J},
  journal={The Journal of Chemical Physics},
  volume={145},
  number={3},
  pages={034107},
  year={2016},
  publisher={AIP Publishing LLC}
}

@article{verma2014increasing,
  title={Increasing the applicability of density functional theory. IV. Consequences of ionization-potential improved exchange-correlation potentials},
  author={Verma, Prakash and Bartlett, Rodney J},
  journal={The Journal of chemical physics},
  volume={140},
  number={18},
  pages={18A534},
  year={2014},
  publisher={American Institute of Physics}
}

@article{verma2016increasing,
  title={Increasing the applicability of density functional theory. V. X-ray absorption spectra with ionization potential corrected exchange and correlation potentials},
  author={Verma, Prakash and Bartlett, Rodney J},
  journal={The Journal of chemical physics},
  volume={145},
  number={3},
  pages={034108},
  year={2016},
  publisher={AIP Publishing LLC}
}

@article{abragam1955hyperfine,
  title={On the hyperfine structure of paramagnetic resonance: the s-electron effect},
  author={Abragam, Anatole and Horowitz, J and Pryce, Maurice Henry Lecorney},
  journal={Proceedings of the Royal Society of London. Series A. Mathematical and Physical Sciences},
  volume={230},
  number={1181},
  pages={169--187},
  year={1955},
  publisher={The Royal Society London}
}

@book{abragam2012electron,
  title={Electron paramagnetic resonance of transition ions},
  author={Abragam, Anatole and Bleaney, Brebis},
  year={2012},
  publisher={Oxford University Press}
}

@article{saue2002four,
  title={Four-component relativistic Kohn--Sham theory},
  author={Saue, Trond and Helgaker, Trygve},
  journal={Journal of computational chemistry},
  volume={23},
  number={8},
  pages={814--823},
  year={2002},
  publisher={Wiley Online Library}
}

@article{spada2022spin,
  title={Spin-density calculation via the graphical unitary group approach},
  author={Spada, Rene FK and Franco, Maur{\'\i}cio P and Nieman, Reed and Aquino, Adelia JA and Shepard, Ron and Plasser, Felix and Lischka, Hans},
  journal={Molecular Physics},
  pages={e2091049},
  year={2022},
  publisher={Taylor \& Francis}
}

@article{nwchem,
author = {Aprà,E.  and Bylaska,E. J.  and de Jong,W. A.  and Govind,N.  and Kowalski,K.  and Straatsma,T. P.  and Valiev,M.  and van Dam,H. J. J.  and Alexeev,Y.  and Anchell,J.  and Anisimov,V.  and Aquino,F. W.  and Atta-Fynn,R.  and Autschbach,J.  and Bauman,N. P.  and Becca,J. C.  and Bernholdt,D. E.  and Bhaskaran-Nair,K.  and Bogatko,S.  and Borowski,P.  and Boschen,J.  and Brabec,J.  and Bruner,A.  and Cauët,E.  and Chen,Y.  and Chuev,G. N.  and Cramer,C. J.  and Daily,J.  and Deegan,M. J. O.  and Dunning,T. H.  and Dupuis,M.  and Dyall,K. G.  and Fann,G. I.  and Fischer,S. A.  and Fonari,A.  and Früchtl,H.  and Gagliardi,L.  and Garza,J.  and Gawande,N.  and Ghosh,S.  and Glaesemann,K.  and Götz,A. W.  and Hammond,J.  and Helms,V.  and Hermes,E. D.  and Hirao,K.  and Hirata,S.  and Jacquelin,M.  and Jensen,L.  and Johnson,B. G.  and Jónsson,H.  and Kendall,R. A.  and Klemm,M.  and Kobayashi,R.  and Konkov,V.  and Krishnamoorthy,S.  and Krishnan,M.  and Lin,Z.  and Lins,R. D.  and Littlefield,R. J.  and Logsdail,A. J.  and Lopata,K.  and Ma,W.  and Marenich,A. V.  and Martin del Campo,J.  and Mejia-Rodriguez,D.  and Moore,J. E.  and Mullin,J. M.  and Nakajima,T.  and Nascimento,D. R.  and Nichols,J. A.  and Nichols,P. J.  and Nieplocha,J.  and Otero-de-la-Roza,A.  and Palmer,B.  and Panyala,A.  and Pirojsirikul,T.  and Peng,B.  and Peverati,R.  and Pittner,J.  and Pollack,L.  and Richard,R. M.  and Sadayappan,P.  and Schatz,G. C.  and Shelton,W. A.  and Silverstein,D. W.  and Smith,D. M. A.  and Soares,T. A.  and Song,D.  and Swart,M.  and Taylor,H. L.  and Thomas,G. S.  and Tipparaju,V.  and Truhlar,D. G.  and Tsemekhman,K.  and Van Voorhis,T.  and Vázquez-Mayagoitia,Á.  and Verma,P.  and Villa,O.  and Vishnu,A.  and Vogiatzis,K. D.  and Wang,D.  and Weare,J. H.  and Williamson,M. J.  and Windus,T. L.  and Woliński,K.  and Wong,A. T.  and Wu,Q.  and Yang,C.  and Yu,Q.  and Zacharias,M.  and Zhang,Z.  and Zhao,Y.  and Harrison,R. J. },
title = {NWChem: Past, present, and future},
journal = {The Journal of Chemical Physics},
volume = {152},
number = {18},
pages = {184102},
year = {2020},
doi = {10.1063/5.0004997}
}

@article{perera2020advanced,
  title={Advanced concepts in electronic structure (ACES) software programs},
  author={Perera, Ajith and Bartlett, Rodney J and Sanders, Beverly A and Lotrich, Victor F and Byrd, Jason N},
  journal={The Journal of Chemical Physics},
  volume={152},
  number={18},
  pages={184105},
  year={2020},
  publisher={AIP Publishing LLC}
}

@article{provasi2001effect,
  title={The effect of lone pairs and electronegativity on the indirect nuclear spin--spin coupling constants in CH 2 X (X= CH 2, NH, O, S): Ab initio calculations using optimized contracted basis sets},
  author={Provasi, Patricio F and Aucar, Gustavo A and Sauer, Stephan PA},
  journal={The Journal of Chemical Physics},
  volume={115},
  number={3},
  pages={1324--1334},
  year={2001},
  publisher={American Institute of Physics}
}

@article{provasi2010optimized,
  title={Optimized basis sets for the calculation of indirect nuclear spin-spin coupling constants involving the atoms B, Al, Si, P, and Cl},
  author={Provasi, Patricio F and Sauer, Stephan PA},
  journal={The Journal of chemical physics},
  volume={133},
  number={5},
  pages={054308},
  year={2010},
  publisher={American Institute of Physics}
}

@article{pou1995density,
  title={Density matrix averaged atomic natural orbital (ANO) basis sets for correlated molecular wave functions: III. First row transition metal atoms},
  author={Pou-Am{\'e}rigo, Rosendo and Merch{\'a}n, Manuela and Nebot-Gil, Ignacio and Widmark, Per-Olof and Roos, Bj{\"o}rn O},
  journal={Theoretica chimica acta},
  volume={92},
  pages={149--181},
  year={1995},
  publisher={Springer}
}

@article{widmark1990malmqvist,
  title={Malmqvist and BO Roos, T heor. Chim. P},
  author={Widmark, PO},
  journal={A{\'e}. Acta},
  volume={77},
  pages={291},
  year={1990}
}

@article{haiduke2018qtp2,
    author = {Haiduke, Roberto Luiz A. and Bartlett, Rodney J.},
    title = "{Non-empirical exchange-correlation parameterizations based on exact conditions from correlated orbital theory}",
    journal = {The Journal of Chemical Physics},
    volume = {148},
    number = {18},
    year = {2018},
    month = {05},
    issn = {0021-9606},
    doi = {10.1063/1.5025723},
    url = {https://doi.org/10.1063/1.5025723},
    pages={184106}
}

@article{widmark1991density,
  title={Density matrix averaged atomic natural orbital (ANO) basis sets for correlated molecular wave functions: II. Second row atoms},
  author={Widmark, Per-Olof and Persson, B Joakim and Roos, Bj{\"o}rn O},
  journal={Theoretica chimica acta},
  volume={79},
  pages={419--432},
  year={1991},
  publisher={Springer}
}

@article{hartree1928, 
title={The Wave Mechanics of an Atom with a Non-Coulomb Central Field. Part I. Theory and Methods}, 
volume={24}, 
DOI={10.1017/S0305004100011919}, 
number={1}, 
journal={Mathematical Proceedings of the Cambridge Philosophical Society}, 
publisher={Cambridge University Press}, 
author={Hartree, D. R.}, 
year={1928}, 
pages={89–110}
}

@article{fock1930,
  title={N{\"a}herungsmethode zur L{\"o}sung des quantenmechanischen Mehrk{\"o}rperproblems},
  author={Fock, Vladimir},
  journal={Zeitschrift f{\"u}r Physik},
  volume={61},
  pages={126--148},
  year={1930},
  publisher={Springer}
}

@article{slater1930note,
  title={Note on Hartree's method},
  author={Slater, John C},
  journal={Physical Review},
  volume={35},
  number={2},
  pages={210},
  year={1930},
  publisher={APS}
}

@article{Slater1972,
  title = {Self-Consistent-Field $X\ensuremath{\alpha}$ Cluster Method for Polyatomic Molecules and Solids},
  author = {Slater, J. C. and Johnson, K. H.},
  journal = {Phys. Rev. B},
  volume = {5},
  issue = {3},
  pages = {844--853},
  numpages = {0},
  year = {1972},
  month = {Feb},
  publisher = {American Physical Society},
  doi = {10.1103/PhysRevB.5.844},
  url = {https://link.aps.org/doi/10.1103/PhysRevB.5.844}
}

@article{vwn,
author = {Vosko, S. H. and Wilk, L. and Nusair, M.},
title = {Accurate spin-dependent electron liquid correlation energies for local spin density calculations: a critical analysis},
journal = {Canadian Journal of Physics},
volume = {58},
number = {8},
pages = {1200-1211},
year = {1980},
doi = {10.1139/p80-159},

URL = { 
    
        https://doi.org/10.1139/p80-159
    
    

},
eprint = { 
    
        https://doi.org/10.1139/p80-159
    
}}

@article{becke1988,
  title={Density-functional exchange-energy approximation with correct asymptotic behavior},
  author={Becke, Axel D},
  journal={Physical review A},
  volume={38},
  number={6},
  pages={3098},
  year={1988},
  publisher={APS}
}

@article{lee1988,
  title={Development of the Colle-Salvetti correlation-energy formula into a functional of the electron density},
  author={Lee, Chengteh and Yang, Weitao and Parr, Robert G},
  journal={Physical review B},
  volume={37},
  number={2},
  pages={785},
  year={1988},
  publisher={APS}
}

@article{Xin2004,
author = {Xin Xu  and William A. Goddard },
title = {The X3LYP extended density functional for accurate descriptions of nonbond interactions, spin states, and thermochemical properties},
journal = {Proceedings of the National Academy of Sciences},
volume = {101},
number = {9},
pages = {2673-2677},
year = {2004},
doi = {10.1073/pnas.0308730100},
URL = {https://www.pnas.org/doi/abs/10.1073/pnas.0308730100},
}

@article{handy2001left,
  title={Left-right correlation energy},
  author={Handy, Nicholas C and Cohen, Aron J},
  journal={Molecular Physics},
  volume={99},
  number={5},
  pages={403--412},
  year={2001},
  publisher={Taylor \& Francis}
}

@article{Cohen2021,
author = { ARON J.   COHEN  and  NICHOLAS C.   HANDY },
title = {Dynamic correlation},
journal = {Molecular Physics},
volume = {99},
number = {7},
pages = {607-615},
year  = {2001},
publisher = {Taylor & Francis},
doi = {10.1080/00268970010023435},
URL = {https://doi.org/10.1080/00268970010023435},
eprint = {https://doi.org/10.1080/00268970010023435}
}

@article{becke1993,
    author = {Becke, Axel D.},
    title = "{Density‐functional thermochemistry. III. The role of exact exchange}",
    journal = {The Journal of Chemical Physics},
    volume = {98},
    number = {7},
    pages = {5648-5652},
    year = {1993},
    month = {04},
    issn = {0021-9606},
    doi = {10.1063/1.464913},
    url = {https://doi.org/10.1063/1.464913},
    eprint = {https://pubs.aip.org/aip/jcp/article-pdf/98/7/5648/11091662/5648\_1\_online.pdf},
}

@article{Stephens1994,
author = {Stephens, P. J. and Devlin, F. J. and Chabalowski, C. F. and Frisch, M. J.},
title = {Ab Initio Calculation of Vibrational Absorption and Circular Dichroism Spectra Using Density Functional Force Fields},
journal = {The Journal of Physical Chemistry},
volume = {98},
number = {45},
pages = {11623-11627},
year = {1994},
doi = {10.1021/j100096a001},
URL = {https://doi.org/10.1021/j100096a001},
eprint = {https://doi.org/10.1021/j100096a001}
}

@article{Adamo1999,
    author = {Adamo, Carlo and Barone, Vincenzo},
    title = "{Toward reliable density functional methods without adjustable parameters: The PBE0 model}",
    journal = {The Journal of Chemical Physics},
    volume = {110},
    number = {13},
    pages = {6158-6170},
    year = {1999},
    month = {04},
    issn = {0021-9606},
    doi = {10.1063/1.478522},
    url = {https://doi.org/10.1063/1.478522},
    eprint = {https://pubs.aip.org/aip/jcp/article-pdf/110/13/6158/10797469/6158\_1\_online.pdf},
}

@article{Zhao2006,
    author = {Zhao, Yan and Truhlar, Donald G.},
    title = "{A new local density functional for main-group thermochemistry, transition metal bonding, thermochemical kinetics, and noncovalent interactions}",
    journal = {The Journal of Chemical Physics},
    volume = {125},
    number = {19},
    year = {2006},
    month = {11},
    issn = {0021-9606},
    doi = {10.1063/1.2370993},
    url = {https://doi.org/10.1063/1.2370993},
    note = {194101},
    eprint = {https://pubs.aip.org/aip/jcp/article-pdf/doi/10.1063/1.2370993/15391406/194101\_1\_online.pdf},
}

@article{Zhao2008,
  title={The M06 suite of density functionals for main group thermochemistry, thermochemical kinetics, noncovalent interactions, excited states, and transition elements: two new functionals and systematic testing of four M06-class functionals and 12 other functionals},
  author={Zhao, Yan and Truhlar, Donald G},
  journal={Theoretical chemistry accounts},
  volume={120},
  pages={215--241},
  year={2008},
  publisher={Springer}
}

@article{grimme2005accurate,
  title={Accurate calculation of the heats of formation for large main group compounds with spin-component scaled MP2 methods},
  author={Grimme, Stefan},
  journal={The Journal of Physical Chemistry A},
  volume={109},
  number={13},
  pages={3067--3077},
  year={2005},
  publisher={ACS Publications}
}

@article{staroverov2003,
  title={Comparative assessment of a new nonempirical density functional: Molecules and hydrogen-bonded complexes},
  author={Staroverov, Viktor N and Scuseria, Gustavo E and Tao, Jianmin and Perdew, John P},
  journal={The Journal of chemical physics},
  volume={119},
  number={23},
  pages={12129--12137},
  year={2003},
  publisher={American Institute of Physics}
}

@article{heyd2003hybrid,
  title={Hybrid functionals based on a screened Coulomb potential},
  author={Heyd, Jochen and Scuseria, Gustavo E and Ernzerhof, Matthias},
  journal={The Journal of chemical physics},
  volume={118},
  number={18},
  pages={8207--8215},
  year={2003},
  publisher={American Institute of Physics}
}

@article{krukau2006influence,
  title={Influence of the exchange screening parameter on the performance of screened hybrid functionals},
  author={Krukau, Aliaksandr V and Vydrov, Oleg A and Izmaylov, Artur F and Scuseria, Gustavo E},
  journal={The Journal of chemical physics},
  volume={125},
  number={22},
  pages={224106},
  year={2006},
  publisher={American Institute of Physics}
}

@article{iikura2001long,
  title={A long-range correction scheme for generalized-gradient-approximation exchange functionals},
  author={Iikura, Hisayoshi and Tsuneda, Takao and Yanai, Takeshi and Hirao, Kimihiko},
  journal={The Journal of Chemical Physics},
  volume={115},
  number={8},
  pages={3540--3544},
  year={2001},
  publisher={American Institute of Physics}
}

@article{datta2015communication,
  title={Communication: Spin densities within a unitary group based spin-adapted open-shell coupled-cluster theory: Analytic evaluation of isotropic hyperfine-coupling constants for the combinatoric open-shell coupled-cluster scheme},
  author={Datta, Dipayan and Gauss, J{\"u}rgen},
  journal={The Journal of Chemical Physics},
  volume={143},
  number={1},
  pages={011101},
  year={2015},
  publisher={AIP Publishing LLC}
}

@article{bergstrom1988radiative,
  title={Radiative lifetime and hyperfine-structure studies on laser-evaporated boron},
  author={Bergstr{\"o}m, H and Faris, GW and Hallstadius, H and Lundberg, Hans and Persson, A and Wahlstr{\"o}m, C -G},
  journal={Zeitschrift f{\"u}r Physik D Atoms, Molecules and Clusters},
  volume={8},
  pages={17--23},
  year={1988},
  publisher={Springer}
}

@article{kaupp2010hyperfine,
  title={Hyperfine coupling constants of the nitrogen and phosphorus atoms: A challenge for exact-exchange density-functional and post-Hartree--Fock methods},
  author={Kaupp, Martin and Arbuznikov, Alexei V and He{\ss}elmann, Andreas and G{\"o}rling, Andreas},
  journal={The Journal of Chemical Physics},
  volume={132},
  number={18},
  pages={184107},
  year={2010},
  publisher={American Institute of Physics}
}

@article{holloway1958determination,
  title={Determination of the Hyperfine Structure of Atomic Nitrogen by Optical Orientation},
  author={Holloway Jr, WW and Novick, R},
  journal={Physical Review Letters},
  volume={1},
  number={10},
  pages={367},
  year={1958},
  publisher={APS}
}

@article{pendlebury1964hyperfine,
  title={Hyperfine structure measurements in 75As, 31P and 53Cr},
  author={Pendlebury, JM and Smith, KF},
  journal={Proceedings of the Physical Society},
  volume={84},
  number={6},
  pages={849},
  year={1964},
  publisher={IOP Publishing}
}

@article{harvey1972diagonal,
  title={Diagonal and Off-Diagonal Hyperfine Structure in the Ground Multiplets of Boron and Aluminum by the Atomic Beam Method; Magnetic Dipole Radial Parameters},
  author={Harvey, JSM and Evans, L and Lew, H},
  journal={Canadian Journal of Physics},
  volume={50},
  number={15},
  pages={1719--1727},
  year={1972},
  publisher={NRC Research Press Ottawa, Canada}
}

@article{harvey1965hyperfine,
  title={HYPERFINE STRUCTURE IN GROUND MULTIPLETS OF $ sup 17$ O AND $ sup 19$ F},
  author={Harvey, J SM},
  journal={Proc. Roy. Soc.(London), Ser. A},
  volume={285},
  year={1965},
  publisher={Oxford Univ.}
}

@article{kusch1949g,
  title={On the g J Values of the Alkali Atoms},
  author={Kusch, P and Taub, H},
  journal={Physical Review},
  volume={75},
  number={10},
  pages={1477},
  year={1949},
  publisher={APS}
}

@article{arimondo1977experimental,
  title={Experimental determinations of the hyperfine structure in the alkali atoms},
  author={Arimondo, Ennio and Inguscio, M and Violino, P},
  journal={Reviews of Modern Physics},
  volume={49},
  number={1},
  pages={31},
  year={1977},
  publisher={APS}
}

@article{pickering1996measurements,
  title={Measurements of the hyperfine structure of atomic energy levels in Co I},
  author={Pickering, Juliet C},
  journal={The Astrophysical Journal Supplement Series},
  volume={107},
  number={2},
  pages={811},
  year={1996},
  publisher={IOP Publishing}
}

@article{ertmer1976zero,
  title={Zero-field hyperfine structure measurements of the metastable states 3 d 2 4 s 4 F 3/2, 9/2 of 45 Sc using laser-fluorescence atomic-beam-magnetic-resonance technique},
  author={Ertmer, W and Hofer, B},
  journal={Zeitschrift f{\"u}r Physik A Atoms and Nuclei},
  volume={276},
  pages={9--14},
  year={1976},
  publisher={Springer}
}

@article{luther1949hyperfine,
  title={The Hyperfine Structure and Nuclear Moments of the Stable Chlorine Isotopes},
  author={Luther Davis, JR and Feld, Bernard T and Zabel, Carrol W and Zacharias, Jerrold R},
  journal={Physical Review},
  volume={76},
  number={8},
  pages={1076},
  year={1949},
  publisher={APS}
}

@article{randolph1975measurement,
  title={A measurement of the ground state hyperfine splitting for Li-like fluorine ions by nuclear quantum beats},
  author={Randolph, WL and Asher, J and Koen, JW and Rowe, P and Matthias, E},
  journal={Hyperfine Interactions},
  volume={1},
  number={1},
  pages={145--150},
  year={1975},
  publisher={Springer}
}

@article{wineland1983laser,
  title={Laser-fluorescence mass spectroscopy},
  author={Wineland, DJ and Bollinger, JJ and Itano, Wayne M},
  journal={Physical Review Letters},
  volume={50},
  number={9},
  pages={628},
  year={1983},
  publisher={APS}
}

@article{malkin2004scalar,
  title={Scalar relativistic calculations of hyperfine coupling tensors using the Douglas--Kroll--Hess method},
  author={Malkin, Irina and Malkina, Olga L and Malkin, Vladimir G and Kaupp, Martin},
  journal={Chemical physics letters},
  volume={396},
  number={4-6},
  pages={268--276},
  year={2004},
  publisher={Elsevier}
}

@article{bahramy2006first,
  title={First-principles calculations of hyperfine parameters with the all-electron mixed-basis method},
  author={Bahramy, MS and Sluiter, MHF and Kawazoe, Y},
  journal={Physical Review B},
  volume={73},
  number={4},
  pages={045111},
  year={2006},
  publisher={APS}
}

@InProceedings{iglo3,
author="Kutzelnigg, Werner
and Fleischer, Ulrich
and Schindler, Michael",
title="The IGLO-Method: Ab-initio Calculation and Interpretation of NMR Chemical Shifts and Magnetic Susceptibilities",
booktitle="Deuterium and Shift Calculation",
year="1991",
publisher="Springer Berlin Heidelberg",
address="Berlin, Heidelberg",
pages="165--262",
isbn="978-3-642-75932-1"
}

@article{aniso-bench,
  title={How accurate is density functional theory in predicting spin density? An insight from the prediction of hyperfine coupling constants},
  author={Witwicki, Maciej and Walencik, Paulina K and Jezierska, Julia},
  journal={Journal of Molecular Modeling},
  volume={26},
  number={1},
  pages={10},
  year={2020},
  publisher={Springer},
  doi={10.1007/s00894-019-4268-0},
  url={https://doi.org/10.1007/s00894-019-4268-0}
}

@article{Mendes_2026,
    author = {Mendes, Rodrigo A. and Windom, Zachary W. and Haiduke, Roberto L. A. and Bartlett, Rodney J.},
    title = {Does correlated orbital theory improve PBE-like functionals?},
    journal = {The Journal of Chemical Physics},
    volume = {164},
    number = {5},
    pages = {054113},
    year = {2026},
    month = {02},
    issn = {0021-9606},
    doi = {10.1063/5.0298139},
    url = {https://doi.org/10.1063/5.0298139},
    eprint = {https://pubs.aip.org/aip/jcp/article-pdf/doi/10.1063/5.0298139/20893086/054113_1_5.0298139.pdf},
}

@article{Mendes_2025,
    author = {Mendes, Rodrigo A. and Windom, Zachary W. and Kim, Hyunsik and Bartlett, Rodney J.},
    title = {On the performance of QTP functionals applied to second-order response properties},
    journal = {The Journal of Chemical Physics},
    volume = {162},
    number = {5},
    pages = {054105},
    year = {2025},
    month = {02},
    issn = {0021-9606},
    doi = {10.1063/5.0246471},
    url = {https://doi.org/10.1063/5.0246471},
    eprint = {https://pubs.aip.org/aip/jcp/article-pdf/doi/10.1063/5.0246471/20370263/054105_1_5.0246471.pdf},
}

@article{Kim_2025,
    author = {Kim, Hyunsik and Perera, Ajith and Mendes, Rodrigo A. and Bartlett, Rodney J.},
    title = {Benchmarking ionization potentials and electron affinities of potential photovoltaic molecules using DFT/QTP functionals and EOM-CC},
    journal = {The Journal of Chemical Physics},
    volume = {163},
    number = {17},
    pages = {174703},
    year = {2025},
    month = {11},
    issn = {0021-9606},
    doi = {10.1063/5.0293131},
    url = {https://doi.org/10.1063/5.0293131},
    eprint = {https://pubs.aip.org/aip/jcp/article-pdf/doi/10.1063/5.0293131/20786381/174703_1_5.0293131.pdf},
}

@misc{Russell_2018,
  title={Computational Chemistry Comparison and Benchmark Database},
  author={Russell III, DJ},
  year={2018},
  publisher={NIST},
  doi = {10.18434/T47C7Z}
}

@article{pc4,
    author = {Jensen, Frank},
    title = {Polarization Consistent Basis Sets. 4: The Elements He, Li, Be, B, Ne, Na, Mg, Al, and Ar},
    journal = {The Journal of Physical Chemistry A},
    volume = {111},
    number = {44},
    pages = {11198-11204},
    year = {2007},
    month = {04},
    issn = {1089-5639},
    doi = {10.1021/jp068677h},
    url = {https://doi.org/10.1021/jp068677h},
    eprint = {https://pubs.acs.org/jpcafh/article-pdf/111/44/11198/39584816/jp068677h.pdf},
}

\end{document}